\documentclass[twocolumn]{aastex631}

\usepackage{natbib}
\usepackage{amsmath}
\usepackage{xspace}
\usepackage{tabularx}
\newcolumntype{Y}{>{\centering\arraybackslash}X}
\usepackage{multirow}
\usepackage{graphicx}
\usepackage{animate}
\usepackage{hyperref}
\usepackage{ocgx2}
\usepackage{threeparttable}  

\def\feii{[Fe\hspace*{1mm}\textsc{II}]~}

\providecommand{\e}[1]{\ensuremath{\times 10^{#1}}}	            

\received{31 December 2021}
\revised{}
\accepted{}

\shorttitle{JWST/NIRCam NGC 4258}
\shortauthors{Fischer et al.}

\begin{document}

\title{JWST NIRCam Imaging of NGC 4258: I. Observation Overview}

\correspondingauthor{Travis C. Fischer}
\email{tfischer@stsci.edu}

\author[0000-0002-3365-8875]{Travis Fischer}
\affil{AURA for ESA, Space Telescope Science Institute, 3700 San Martin Drive, Baltimore, MD 21218, USA}
\author{Nicholas F. Cothard}
\affil{NASA Goddard Space Flight Center, 8800 Greenbelt Road, Greenbelt, MD, USA}
\author{Omnarayani Nayak}
\affil{NASA Goddard Space Flight Center, 8800 Greenbelt Road, Greenbelt, MD, USA}
\author{Henrique Schmitt}
\affil{Naval Research Laboratory, 4555 Overlook Ave. NW, Washington DC-20375, USA}
\author{Erin Smith}
\affil{NASA Goddard Space Flight Center, 8800 Greenbelt Road, Greenbelt, MD, USA}
\author{Jason Glenn}
\affil{NASA Goddard Space Flight Center, 8800 Greenbelt Road, Greenbelt, MD, USA}




\begin{abstract}
\noindent

We present James Webb Space Telescope (JWST) NIRCam imaging of the nearby Seyfert 1.9 galaxy NGC 4258, which hosts strong star formation regions as well as an anomalous jet-like radio structure that extends through a significant portion of its disk. This galaxy provides a unique environment to study Active Galactic Nucleus (AGN)-driven shocks and their impact on the interstellar medium (ISM) as its proximity allows for narrow-band observations of various near-infrared tracers sensitive to multiple levels of shock and radiative excitation: \feii (1.64 $\mu$m), Pa$\alpha$ (1.87 $\mu$m), H$_2$ (2.21 $\mu$m), 3.3 $\mu$m polycyclic aromatic hydrocarbon (PAH) emission, Br$\alpha$ (4.05 $\mu$m), and Pf$\beta$ (4.66 $\mu$m), allowing us to trace shocks with parsec-scale resolution. Comparing these near-infrared observations with available ultraviolet, optical, radio, and X-ray imaging, we find that shocks present in the brightest regions of the anomalous radio structure are likely of low-velocity (50-100 km s$^{-1}$), suggesting that these features originate from AGN-driven winds that interact with the host medium and mechanically impart energy into the disk. Further, while co-spatial \feii and H$_2$ emission indicate multi-phase shocks, PAH emission is relatively weaker or absent in the most shock-excited regions, consistent with the destruction of small dust grains. Finally, we propose that surveys identifying enhanced \feii in AGN host galaxies may systematically reveal a key population where AGN feedback is significantly coupled with the surrounding ISM and actively shaping galaxy evolution.
\end{abstract}

\keywords{}


\section{Introduction} 
\label{sec:intro}
\noindent


Shocks, including those generated by outflows from Active Galactic Nuclei (AGN), can significantly influence the ISM in galaxies, with diverse effects observed under various conditions. Depending on the shock strength, gas density, and galaxy environment, shocks have been observed or modeled to heat and compress interstellar gas  \citep{All08}, affect molecular composition and ionization states \citep{Ala13,Vit14,Mee22}, ablate dust grains \citep{Gui11}, inject turbulent energy into the ISM  \citep{Sot24}, and, in some cases, can even trigger gravitational collapse and star formation \citep{Sil13,Zub13}. However, the prominence of these effects varies widely among galaxies, depending strongly on local physical conditions and the intensity of AGN feedback \citep{Ram22,Siv25}. With $\sim15\times$ better angular resolution (e.g., JWST $3.3~\mu$m versus {\it Spitzer} $8~\mu$m) and substantially improved sensitivity, JWST enables a detailed investigation of the impact of AGN on the interstellar medium (ISM) in nearby galaxies at parsec-scale spatial resolution.

Seyfert AGN NGC 4258 possesses a unique set of characteristics that make it ideal for investigating the impact of an AGN on the ISM using the infrared spectral coverage provided by JWST imaging including a well-studied nucleus, a gas- and dust-rich galactic disk, and an anomalous 'jet' that intersects the disk gas. The nucleus houses a $3.8\pm0.3\times10^7$ M$_{\odot}$ black hole \citep{Sag16}, with an infrared ($1-20~\mu$m) luminosity of $2\times10^8$ L$_{\odot}$ driven by an accretion rate of approximately $10^{-3}$ M$_{\odot}$ yr$^{-1}$ \citep{Cha00}. It is this accretion that has been thought to fuel the anomalous 'jet' — termed anomalous because it partially traverses the plane of the disk and has resulted in strongly shock-excited H$_2$, CO, and [C~II] emission \citep{ogle_jet-shocked_2014,appleton_jet-related_2018}. Previous {\it Spitzer} $8~\mu$m observations were employed to seek evidence of dust destruction due to jet-induced shocks \citep{Lai10}. However, no evidence was discovered, leading the authors to conclude that a likely $10\times$ improvement in angular resolution would be necessary to measure the effects of the jet shocks on the interstellar dust. 

NGC 4258 has a precise, megamaser-derived distance of $7.586\pm0.112$ Mpc and has served as a key rung on the cosmic distance ladder \citep{Her99}. It is among the nearest galaxies hosting a well-studied AGN; for context, several others, such as NGC 4945 ($\sim$3.4 Mpc), Centaurus A ($\sim$3.8 Mpc), and NGC 4395 ($\sim$4.3 Mpc), lie closer. NGC 4258's close proximity to the Milky Way is critical for these observations, as we can utilize the narrow- and medium-band filters of NIRCam aboard JWST to isolate several important emission features. In contrast, most other AGN are at much greater redshifts and thus require integral field unit (IFU) spectroscopy for similar analyses, which occurs at a much smaller scale and with lower spatial resolution and sensitivity. Therefore, this case study acts as a crucial testbed for examining shock-driven processes in galaxies, which is necessary for understanding the physics of high-redshift galaxies where parsec-scale resolution is not feasible.


\begin{table}[h!]
\caption{NGC 4258 (M106) Properties}
\label{tab:ngc4258}
\footnotesize
\begin{threeparttable}
\begin{tabular}{ll}
\hline
Parameter \& Value \\
\hline\hline
RA \& Dec (J2000)\tnote{a} & 12$^{\rm h}$18$^{\rm m}$58$^{\rm s}$,\, 47$^\circ$18'14'' \\
Classification\tnote{b}          & Seyfert 1.9, SAB(s)bc  \\
log M$_{BH}$\tnote{c} \, ($\mathrm{M}_{\odot}$) & 7.58 \\  
log L$_{1-20 \mu m}$\tnote{d} \, (erg s$^{-1}$)  &  41.88  \\
log L$_{2-10\,keV}$\tnote{e} \, (erg s$^{-1}$) & 41.0 \\
Distance\tnote{f} \, (Mpc) & $7.576\pm0.112$  \\ 
Angular scale & 36.74 pc/arcsec \\
\hline
\end{tabular}
\begin{tablenotes}
\item[a] \cite{Herrnstein_2005}
\item[b] \cite{Veron_2006}
\item[c] \cite{Sag16}
\item[d] \cite{Cha00}
\item[e] \cite{Mas12}
\item[f] \cite{Reid_2019}
\end{tablenotes}
\end{threeparttable}
\end{table}

\begin{figure*}
\centering
\includegraphics[width=\textwidth]{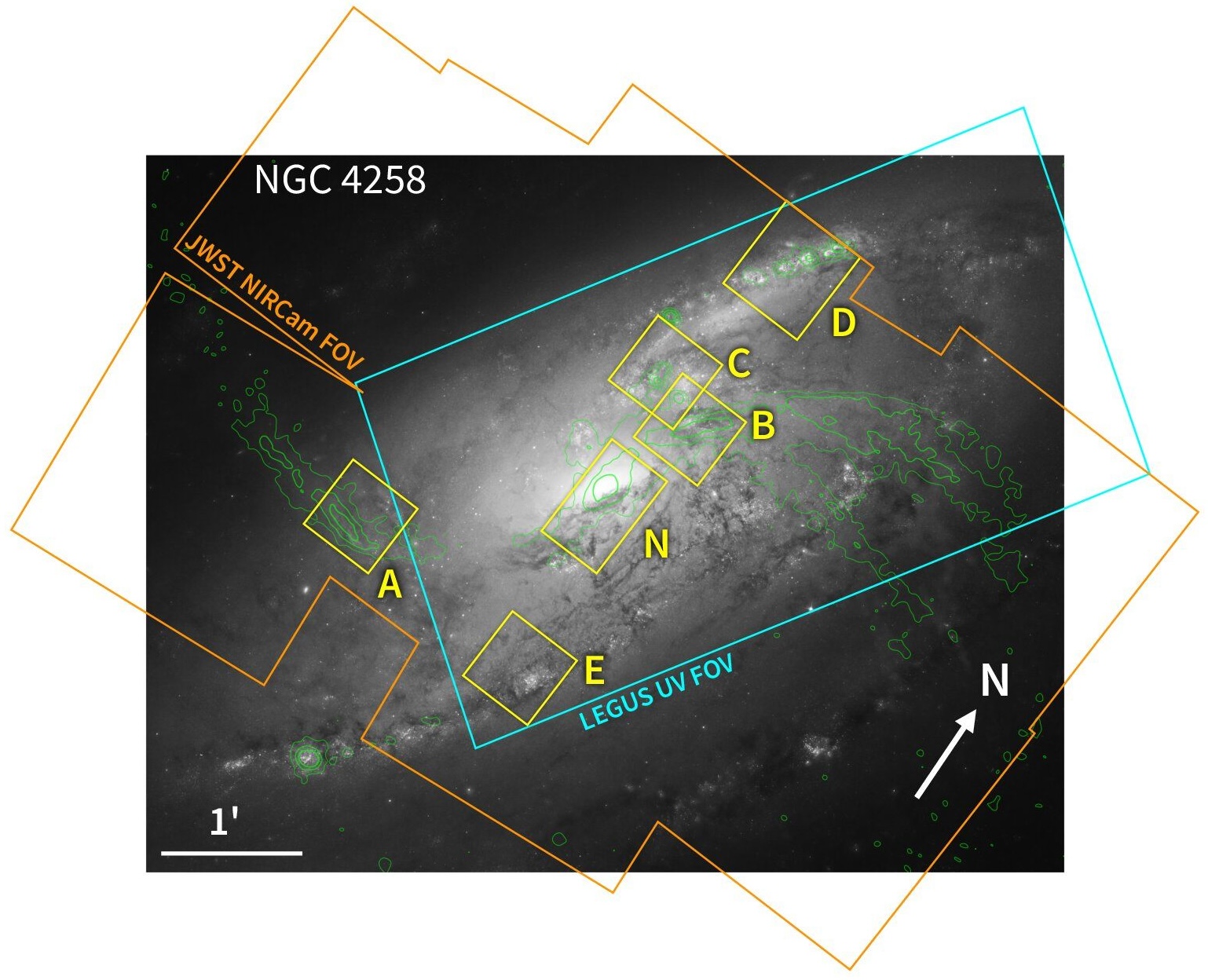}

\caption{Eight-filter composite optical imaging of NGC 4258. The orange outline depicts the approximate footprint of the NIRCam long wavelength channel field of view. The cyan outline depicts the footprint of archival HST WFC3/F275W imaging from the LEGUS survey \citep{Cal15}. Green contours represent VLA 4.89 GHz radio continuum emission from the NRAO/VLA Archive Survey (NVAS; image credit: NRAO/VLA Archive Survey, \textcopyright{} 2005--2009 AUI/NRAO)} Yellow boxes highlight several regions of interest detailed in Figures~\ref{fig:grid_nucleus}--\ref{fig:grid_E}.
\label{fig:fov}
\end{figure*}

To understand the role of shocks in NGC 4258, we focus on several critical emission-line diagnostics. The \feii 1.6 $\mu$m line is frequently used as a shock tracer because interstellar iron, typically locked within dust grains, is liberated into the gas phase through shock-induced sputtering \citep{Gui11,All08}. Once freed, iron atoms can be singly ionized (\feii) by relatively modest radiation fields, given their ionization potential of 7.9 eV. Consequently, \feii emission often arises in regions of strong shocks and has been widely used to trace shock interactions across a variety of astrophysical environments, including AGN-driven outflows \citep{Rod04,Koo16}, supernova remnants \citep{Koo15}, and protostellar jets \citep{Nis02}. Specifically, prominent \feii emission typically requires shock velocities on the order of 100 km s$^{-1}$ or greater \citep{All08,Koo16}, making it a robust indicator of energetic shock conditions like those expected where the anomalous jet-like radio structure in NGC 4258 impacts the ISM.

Molecular hydrogen (H$_2$) emission complements this picture by tracing lower-velocity shocks, typically defined as those with $v \lesssim 50$ km s$^{-1}$. In such shocks, the kinetic energy is insufficient to dissociate H$_2$, allowing the molecules to survive and emit via rovibrational transitions. These are often C-type shocks, common in molecular gas where magnetic fields and high pre-shock densities moderate the shock front \citep{Hol89,Flo10}. In addition to shocks, H$_2$ can also be excited by non-collisional mechanisms such as ultraviolet fluorescence \citep{Bla87} and X-rays \citep{Mal96}, depending on the local radiation field and gas conditions.

From studies of line ratios, line widths, and morphologies in a sample of 21 Seyfert galaxies observed in H$_2$ and \feii, \citep{Rod05} concluded that nuclear H$_2$ generally arises from X-ray excitation. However, in the case of NGC 4258, \citep{ogle_jet-shocked_2014} concluded that the H$_2$ emission in the disk arises from shocks (or possibly cosmic rays). Hydrogen recombination lines like Pa-$\alpha$ and Br-$\alpha$ provide additional diagnostics of star formation, with Pa-$\alpha$ correlating well with far-infrared luminosity and star formation rates (e.g., \citealt{rieke_determining_2009}). Br-$\alpha$, with its longer wavelength, experiences significantly lower extinction than Pa-$\alpha$—by a factor of about 5 in terms of $A_\lambda$ (e.g., \citealt{rieke_extinction_1985}). This makes Br-$\alpha$ especially valuable for probing star formation in regions that are deeply embedded in dust, where even Pa-$\alpha$ may be attenuated.

Polycyclic Aromatic Hydrocarbons (PAHs) are likely created in AGB star envelopes and distributed through the interstellar medium by the evolved star winds. They are observed as broad non-Gaussian emission bands observed throughout the near and mid-infrared, including at 3.3, 6.2, 7.7, and $11.3~\mu$m \citep{All89}. PAH emission, which arises from excitation by ultraviolet and optical photons \citep{Draine2021}, is strongly associated with star formation in galaxies \citep{Pee04}. While weaker than the mid-IR PAH bands, the $3.3~\mu$m feature has been shown to vary comparatively little with stellar radiation field \citep{vanD04}, and may be of greater use in observing high-redshift galaxies with JWST due to lower wavelength. While observed in star-forming galaxies, 3.3 $\mu$m emission is not widely detected in AGN, possibly because PAHs are sufficiently fragile to be destroyed by interstellar shocks or intense ionizing radiation fronts \citep{Kim12}. However, recent JWST observations have begun to clarify the behavior of PAH features in AGN hosts, showing that PAH strength and ionization state can vary with distance from the nucleus and with AGN luminosity (e.g., \citealt{Gar22, Gar24, Rig24, Don23, Don24}). In NGC 4258, PAH emission provides a potential lens to examine the impact of AGN feedback on dust and star formation processes.


Observations of these tracers of excited ISM together assess the impact of AGN feedback on the NGC 4258 interstellar medium: \feii and H$_2$ reveal gradients of shock strength, while Pa$\alpha$ and Br$\alpha$ signify star formation, potentially indicating positive feedback. PAH emission highlights ionization fronts, particularly in star-forming regions. Collectively, they examine shock energetics and the ionized, atomic, and molecular phases of the ISM, where dust grains may have been destroyed. One of our investigation's objectives is to identify structure — possibly stratification — in the NGC 4258 ISM resulting from the energy deposition linked to the jet-like radio structure near the nucleus, along the radio-ISM interface, or where the jet-like structure appears to be deflected (Figure \ref{fig:fov}). Similar NIRCam imaging analyses have recently been used to identify stratified ISM structure and dust feedback signatures in nearby galaxies (e.g., \citealt{Cha24}). 

Fully understanding the complex interactions in NGC 4258 not only affords insight into this galaxy itself but can also help interpret data from high-z galaxies. For high-redshift galaxies, the lack of spatial resolution makes separating galaxy elements challenging if not impossible, but by investigating NGC 4258 and applying the findings, we can better interpret the high-redshift universe.

\section{NIRCam Observation Strategy} 
\label{sec:obsstrat}
\begin{figure}
\centering
\includegraphics[width=0.47\textwidth]{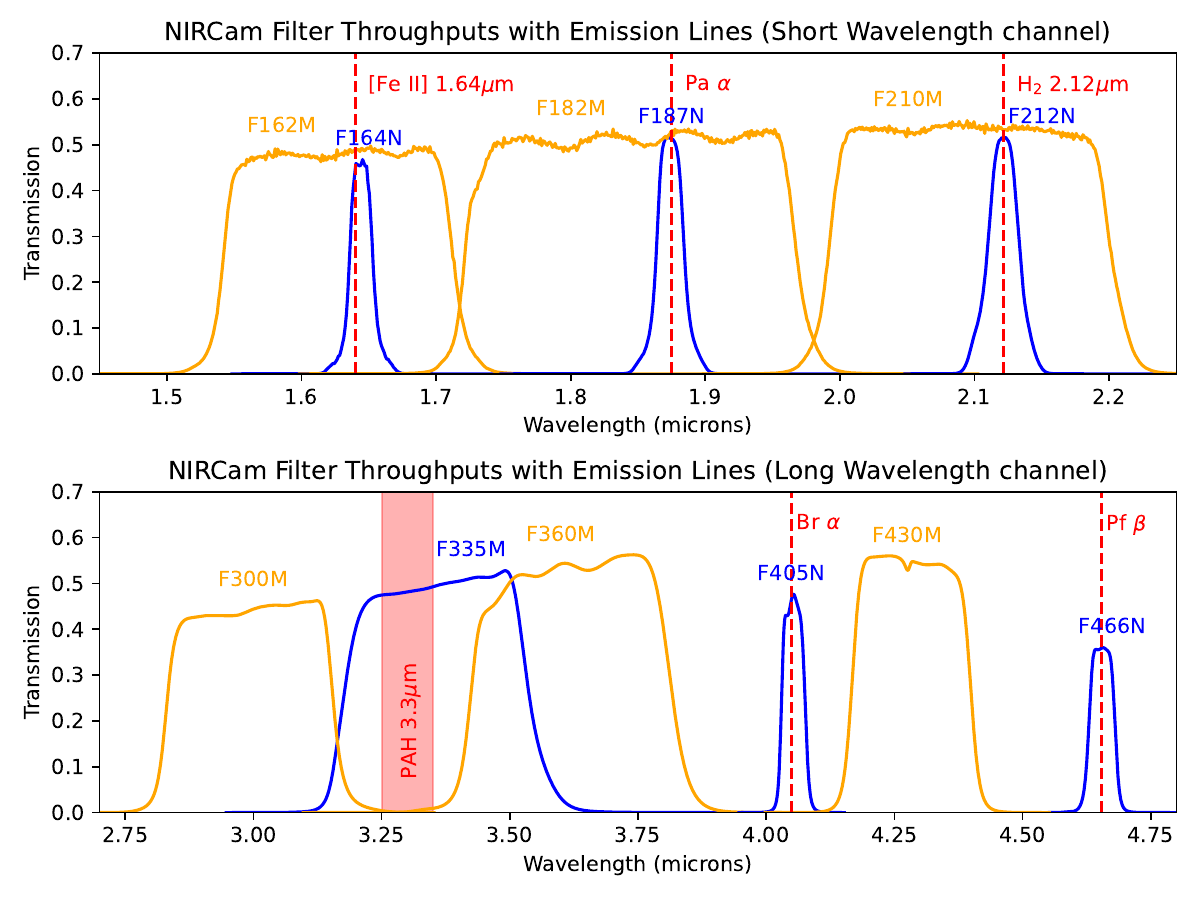}
\caption{NIRCam filter bandpasses, emission lines, and PAH band. Top: Bandpasses for the F164N, F162M, F187N, F182M, F212N, and F210M filters.  Bottom: Bandpasses for the F335M, F300M, F360M, F405N, F466N, and F430M filters.  Locations of emission lines are identified with red dashed lines and the approximate position of the PAH 3.3$\mu$m feature is highlighted as a red, shaded region in the bottom panel.
}
\label{fig:filters}
\end{figure}

\noindent
Using the Near-Infrared Camera (NIRCam \cite{rieke_performance_2023}) aboard JWST, we imaged NGC 4258 (PID 2080; PI: J. Glenn). We targeted the central $\approx3.5'$ using a 2-row observing mosaic with 10$\%$ overlap, designed to encompass both the anomalous radio arc and the star-forming arm regions of the galaxy. The NIRCam observations were conducted using all modules and full subarrays. A $2\times1$ mosaic (10$\%$ overlap) was implemented with a \textsc{full3} dither pattern and no subpixel dithers. The observations were performed in two sessions, roughly one year apart. The first set of observations occurred on March 25 and 26, 2023; however, half of the 36 observation visits were skipped due to a guide star acquisition failure. As such, additional visits were scheduled and completed on March 21, 2024. Unfortunately, 7 visits were again skipped because of guide star failures. The resulting mosaic footprint from the successful 29 visits is illustrated in Figure \ref{fig:fov}. 

Observations included simultaneous imaging with the short-wavelength (SW) and long-wavelength (LW) filter combinations of F164N/F405N, F187N/F466N, F212N/F430M, F162M/F300M, F182M/F335M, and F210M/F360M. Filters were chosen to either isolate specific emission lines or perform continuum subtraction, as illustrated in Figure \ref{fig:filters}. In order to optimize the signal-to-noise ratio of the images, different readout patterns and exposure parameters were chosen for the narrow and medium band filter images. The \textsc{medium8} readout pattern with 6 groups per integration and 4 integrations per exposure were used for F164N, F187N, F212N, F405N, F466N, and F430M. The \textsc{shallow4} readout pattern with six groups per integration and 1 integration per exposure were used for F162M, F182M, F210M, F300M, F335M, and F360M. A total exposure time of 7569 seconds was used for SW and LW filter combinations F164N/F405N, F187N/F466N, and F212N/F430M. A total exposure time of 934 seconds was used for SW/LW filter combinations F162M/F300M, F182M/F335M, and F210M/F360M. These times reflect the nominal exposure durations submitted in the observing plan; actual exposure times may be lower in regions where visits were skipped due to guide star acquisition failures.

\section{Data Reduction} 
\label{sec:datared}
\subsection{NIRCam Image Processing}
\label{sec:image_processing_method}

\noindent
We reduced the data using the standard JWST pipeline with some exceptions described here.  We processed NIRCam SW and LW tiles using pipeline version 1.13.4 for all stage 1 and stage 2 \citep{bushouse_2023}. We applied slight modifications by using the 1/f noise correction \citep{willott_image1overf_2024}, and JWST Hubble Alignment Tool (JHAT) \citep{rest_jhat_2024} during the stage 3 step of the pipeline processing. We used Operational Pipeline Calibration Reference Data System (CRDS) \texttt{jwst\_1094.pmap} for reducing the data. We used JHAT to align all images to the F430M mosaic since this is the longest LW filter. The final F430M alignment catalog excluded sources from one of the NIRCam tiles due to alignment issues, which could not be resolved and will be discussed in a later paper.  It did not affect the results presented in this paper. Because alignment to F430M was already done with the JHAT step, the stage 3 \textit{tweakreg} step was skipped. Additionally, we set the pixel scale of 0.03\arcsec\ for all mosaics.




To prepare the mosaics for continuum subtraction, it was important that each narrow and medium band pair of mosaics was reprojected onto a common pixel grid and world coordinate system (WCS). We reprojected each mosaic to the longest-wavelength medium-band filter, F430M, so that each reprojected mosaic could be compared pixel-for-pixel to the others. We used \texttt{reproject\_interp} from the \texttt{reproject} library\footnote{\url{https://reproject.readthedocs.io/en/stable/api/reproject.reproject_interp.html}} for this.

\subsection{Continuum Subtraction}
\label{sec:continuum_subtraction_method}


The narrow-band filter NIRCam images of NGC 4258 are dominated by underlying continuum emission. To isolate the targeted emission features, the continuum contribution must first be removed. For \feii, Pa$\alpha$, and H$_2$, images were acquired with medium-band filters that spectrally encompass the narrow-band filter images. For Br$\alpha$ and Pf$\beta$, spectrally adjacent medium-band filters were used to estimate the continuum contribution. This method is consistent with similar photometric subtraction approaches in the literature \citep[e.g.,][]{Sto18}. Line emission images were created by subtracting a scaled medium-band filter image from the narrow-band filter image, with the scale factor given by the median value of the narrow-to-medium band intensities. Pixels where the signal-to-noise ratio (SNR) of either the narrow- or medium-band filter images was less than 5 were discarded prior to computing the median value. This method assumes that the bandpasses in the majority of the imaged fields are continuum-dominated, which is generally true. The continuum subtraction method works best where the SNR is high. The fainter shock-tracing emission lines present challenges to finding a single scale factor that accurately nulls the emission in both the nucleus and the less-dense regions of the galaxy, resulting in over-subtraction in one of these areas.

The faint surface brightness of the [Fe~II] and H$_2$ emission lines also highlights defects in image processing, such as 1/f noise and sky background mismatches. Significant improvements were achieved by switching from the standard \texttt{jwst} pipeline removal method to the \texttt{image1overf.py} method from \cite{willott_image1overf_2024}. For context, bright \feii arcs associated with the radio hotspots exhibit peak surface brightnesses of 2-4 MJy sr$^{-1}$, while Br$\alpha$ structures adjacent to these arcs typically reach 20-50 MJy sr$^{-1}$. This substantial difference emphasizes the importance of tailored background subtraction and noise mitigation strategies to reliably recover the low-surface-brightness shock emission. Among the \texttt{skymatch} background subtraction options in the \texttt{jwst} \texttt{Image3} pipeline, the ``local'' method provided the most reliable results for our dataset, minimizing over- and under-subtraction across individual tiles. The ``match'' method could not be applied due to insufficient overlap between images, while the ``global'' method resulted in background oversubtraction and artificially low flux levels. The ``local'' method computes the sky value per image using masked statistics (mean, median, and mode), and led to improved surface brightness uniformity across the mosaic. While morphological features were consistent across methods, this choice improved photometric consistency in faint emission regions.

To further validate the continuum subtraction, we analyzed three $35'' \times 35''$ regions across the field. Two of these (Regions A and B in Figure~\ref{fig:fov}) contain significant [FeII] emission, while a third, emission-free region centered at 12$^{\rm h}$18$^{\rm m}$49$\fs$2523 +47$\degr$18$\arcmin$47$\farcs$286 is dominated by continuum. In Regions A and B, 66.4$\%$ and 89.3$\%$ of pixels, respectively, show less than 5$\%$ change after subtraction, indicating that most of the field is indeed continuum dominated. Even in Region C, 72.3$\%$ of pixels meet this criterion. We also evaluated the standard deviation in Region C after continuum subtraction and found a $1\sigma$ residual scatter of 0.10MJysr$^{-1}$. When scaled to the extraction aperture, this corresponds to a photometric uncertainty of $\sim$4Jy. This level of uncertainty is consistent with the [FeII] flux errors reported in Table\ref{tab:table1}, demonstrating that the continuum subtraction process does not introduce significant systematic errors. Representative cutouts and flux histograms for all three regions are provided in Appendix~X to illustrate the subtraction performance across emission- and continuum-dominated areas.

\section{Additional Data}
\label{sec:additional_data}

To gain a deeper understanding of our NIRCam observations regarding the physical processes occurring in NGC 4258, we also retrieved several archival datasets at different wavebands to incorporate into our analysis.

\subsection{Jansky Very Large Array Imaging}

We provide archival imaging of NGC 4258 from the NRAO Jansky Very Large Array (VLA) Archive Survey (NVAS)\footnote{http://archive.nrao.edu/nvas} at 4.89 GHz. C-configuration observations took place on December 16, 1986, with a restoring beam of $\alpha \times \delta = 4.48'' \times 3.88''$ at a position angle of $\theta = -59^{\circ}$, and achieve a root mean square (rms) noise of $\sim$22 $\mu$Jy\,beam$^{-1}$. The field of view of these observations encapsulates those of our JWST observations. 

\subsection{Hubble + Ground-based Optical Imaging}

The optical image of NGC 4258, utilized in our analysis and displayed in Figure \ref{fig:fov}, originated from a press release\footnote{https://hubblesite.org/contents/news-releases/2013/news-2013-06.html} and was produced in collaboration with astrophotographers to address gaps where Hubble Space Telescope (HST) data was either incomplete or unavailable. This mosaic integrates data from several HST instruments, including ACS/WFC, WFPC2, and WFC3/UVIS, alongside ground-based observations. Specifically, data from HST Program 11570 (PI: A. Riess) and additional ACS and WFPC2 datasets were incorporated. The applied filters include F435W (G), F555W (V), F606W (V), and F814W (I) for ACS/WFC, F656N (H$\alpha$) for WFPC2, as well as F555W (V) and F814W (I) for WFC3/UVIS. Ground-based H$\alpha$ data at 656 nm was provided by R. Gendler and J. GaBany. Due to the nature of the image assembly process, a fully merged FITS mosaic was not available; however, a WCS-solved FITS image was generated from the press image. The initial alignment with reference data using the WCS solution was further refined by visually aligning the dust lanes in the optical image with PAH structures in the near-IR. Although this image lacks useful flux information, it is used solely for morphological comparisons between the optical and near-IR data. 

\subsection{Hubble LEGUS UV Imaging}

We also incorporated archival ultraviolet WFC3 F275W ($\lambda$2704\AA) filter HST imaging into our analysis from the Legacy ExtraGalactic UV Survey (LEGUS) Program \citep{Cal15}. Observations occurred 05 and 17 March 2014, and were designed to reach a UV surface brightness of 3.5 $\times$ 10$^{-19}$ erg s$^{-1}$ cm$^{-2}$ $\AA^{-1}$ arcsec$^{-2}$. Figure \ref{fig:fov} illustrates the field of view covered in the program compared to that of our JWST observations. 

\subsection{Chandra X-ray Imaging}

To examine the high-energy emission associated with the physical processes in NGC 4258, we retrieved archival \textit{Chandra} X-ray Observatory imaging taken with the Advanced CCD Imaging Spectrometer (ACIS-S). Observations were conducted on 28 May 2001 (ObsID 1618) in the Timed Exposure mode with the Very Faint telemetry format. The total exposure time was 32 ks. The dataset was obtained from the \textit{Chandra} Data Archive. The ACIS-S instrument provides high spatial resolution ($\sim$0.5'') and sensitivity to X-ray emission from hot gas, shocks, and non-thermal processes, allowing us to investigate the interaction between AGN feedback and the surrounding interstellar medium. 




\section{Results} 
\label{sec:results_discussion}

\begin{figure*}
\centering
\includegraphics[width=\textwidth]
{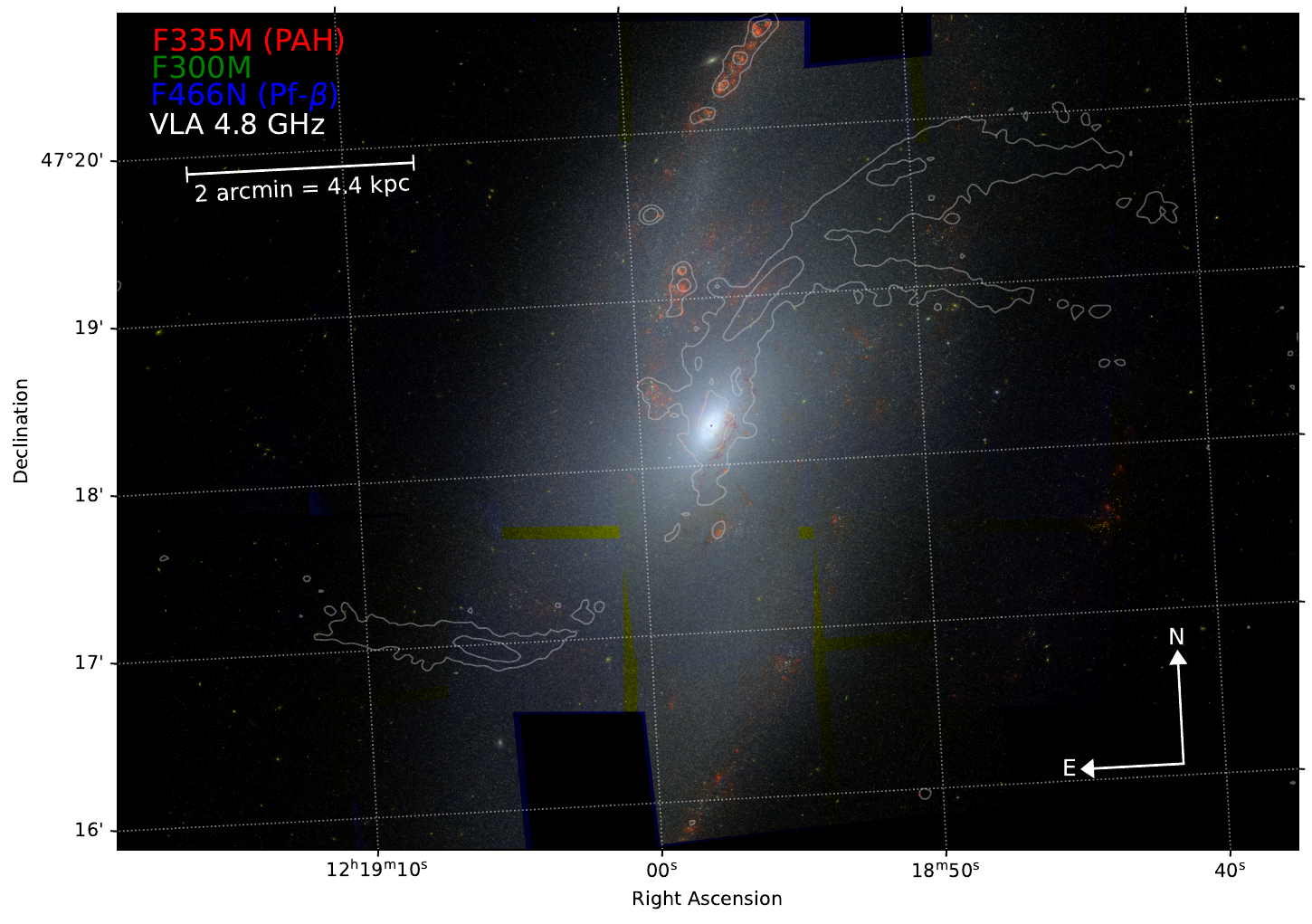}
\caption{Three-color near-infrared composite mosaic of NGC 4258 using NIRCam filters, F466N (blue), F300M (green), and F335M (red). VLA 4.8 GHz images are overplotted as contour lines. Diffuse continuum emission is observed throughout the galaxy. At a distance of 7.2 Mpc, NIRCam resolves star clusters and potentially individually stars. Bands of PAH and Paschen-$\alpha$ emission trace the spiral structure of the galaxy.}
\label{fig:BroadbandMosaic}
\end{figure*}

Figure \ref{fig:BroadbandMosaic} shows a three-color image of NGC 4258 using the F335M (red), F300M (green), and F466N (blue) NIRCam filters. The image largely resembles previous \textit{Hubble} imaging of NGC 4258 in Figure \ref{fig:fov}, but with significantly less extinction from the galaxy's dust lanes, which now only remain visible in small regions near the nucleus at shorter wavelength bands.

At a distance of 7.2 Mpc, the SW NIRCam filter images (before reprojection) have a pixel scale of approximately 1.1 pc ($\sim0.031''$), while the LW NIRCam filter images have a pixel scale of approximately 2.2 pc ($\sim0.063''$). Compared to \textit{Spitzer} images of NGC 4258, this represents an increase in spatial resolution by roughly factors of 10 to 20. Individual stars or star clusters are resolved throughout the galaxy. A study of the observed stellar populations will be explored in depth in a future paper (Nayak et al., in prep).

Although several images using the selected narrow-band filters show excess emission along the northern and southern spiral arms of the galaxy, we perform the continuum subtraction method described in Section \ref{sec:continuum_subtraction_method} to draw out the targeted, faint emission-line features.

\subsection{Continuum Subtracted Emission Line Maps}
\label{sec:line_maps}

\begin{figure*}
\centering
\includegraphics[width=\textwidth]
{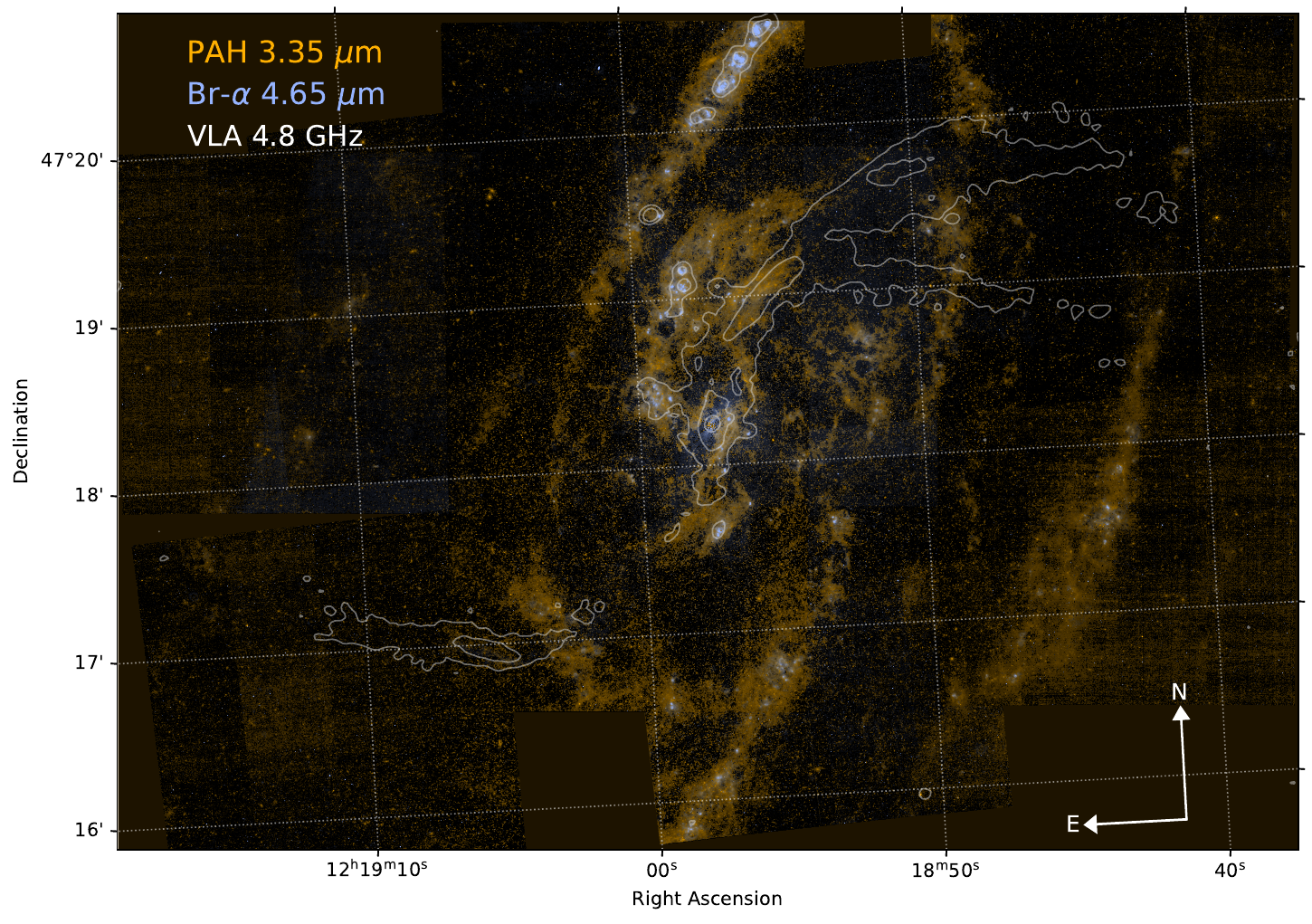}
\caption{Two-color composite mosaics of continuum subtracted PAH (yellow) and Br$-\alpha$ (blue). VLA 4.8 GHz emission is overplotted as a contour line. PAH emission traces dust lanes and outlines the star formation regions.}
\label{fig:ContSubMosaicPAHBr}
\end{figure*}


Comparing morphologies across each continuum-subtracted emission line map, we find several correlations.

Figure \ref{fig:ContSubMosaicPAHBr} highlights comparisons between 3.35 $\mu$m PAH emission and continuum-subtracted Br$\alpha$ emission. Bright PAH 3.3\,$\mu$m emission largely encapsulates star-forming regions traced by the Br$\alpha$ hydrogen emission. As hydrogen recombination lines are produced in the ionized gas surrounding massive stars, commonly referred to as H~II regions, these features correlate well with far-infrared luminosity and star formation rates in galaxies (e.g., \citealt{rieke_determining_2009}). We know these are star-forming regions as they also are exceedingly bright in the ultraviolet (see work from LEGUS program; \citealt{Cal15}). This relationship reveals several examples of stratification in the interstellar medium due to energy deposition in star-forming regions throughout the galaxy, further described in Section \ref{sec:results_discussion}. The revealed PAH structure is similar in morphology to 7.7$\mu$m PAH emission observed in previous \textit{Spitzer} imaging \cite{ogle_jet-shocked_2014}, but with much higher spatial resolution. Generally, the PAH and Br$\alpha$ emission form an off-center ring of $r\sim$ 130 arcsec ($\sim$4.8 kpc), with an exterior spur to the south and an interior knot north of the nucleus. While there appears to be some overlap between radio emission and the interior northern PAH knot, the radio arms largely remain unassociated with the PAH morphology. However, faint, diffuse Br$\alpha$ emission can be seen in the radio arc at the locations of the brightest radio flux.

\begin{figure*}
\centering
\includegraphics[width=\textwidth]
{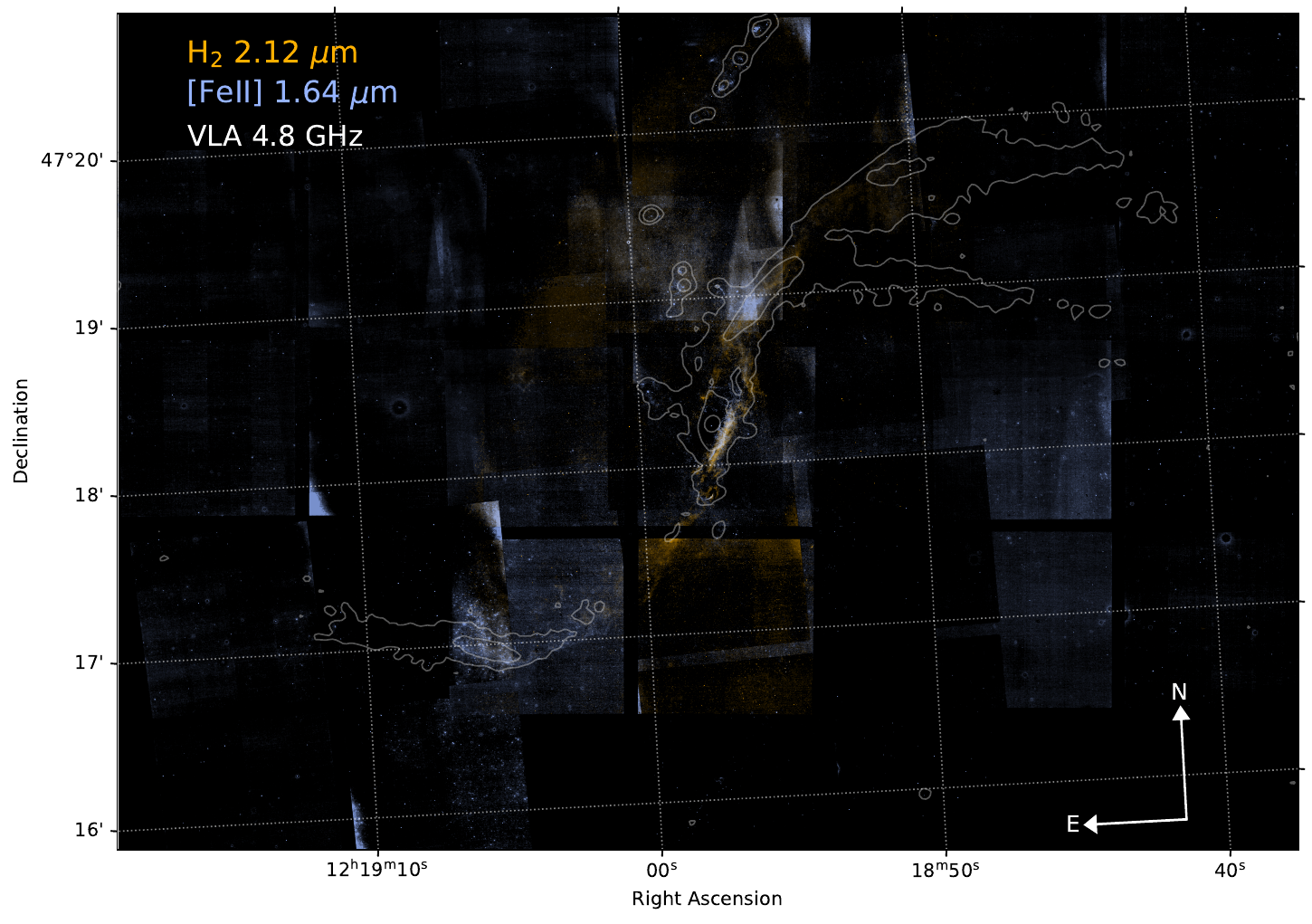}
\caption{Two-color composite mosaics of continuum subtracted [Fe~II] (blue) and H$_2$ (yellow). VLA 4.8 GHz emission is overplotted as a contour line. The shock-tracing \feii and H$_2$ emission lines follow the anomalous radio arc structure rather than the spiral structure of the galaxy. 1/f noise and stray light artifacts are amplified by the continuum subtraction process due to the faint \feii and H$_2$ emission.}
\label{fig:ContSubMosaicFeiiH2}
\end{figure*}

Figure \ref{fig:ContSubMosaicFeiiH2} compares the morphologies of the two shock-tracing emission lines, \feii and H$_2$. Contrasting with the previous map, both lines show a strong correlation with the morphology of the radio continuum. \feii is largely isolated to three regions; adjacent to the nucleus to its southwest and in the brightest knots of both the northern and southern anomalous radio arc. Strong H$_2$ emission generally follows the southeast-to-northwest structure of the radio arcs, connecting regions of nuclear PAH emission surrounding the AGN to the \feii emission in the bright radio knots. 

While these maps both give a global sense of physical processes occurring in NGC 4258, we also provide several 'postage stamp' regions selected to highlight the interplay between emission line morphologies and how they map to the host galaxy as observed in the optical. The distribution of these regions is illustrated in Figure \ref{fig:fov}.

\begin{figure*}
\centering
\includegraphics[width=0.98\textwidth]{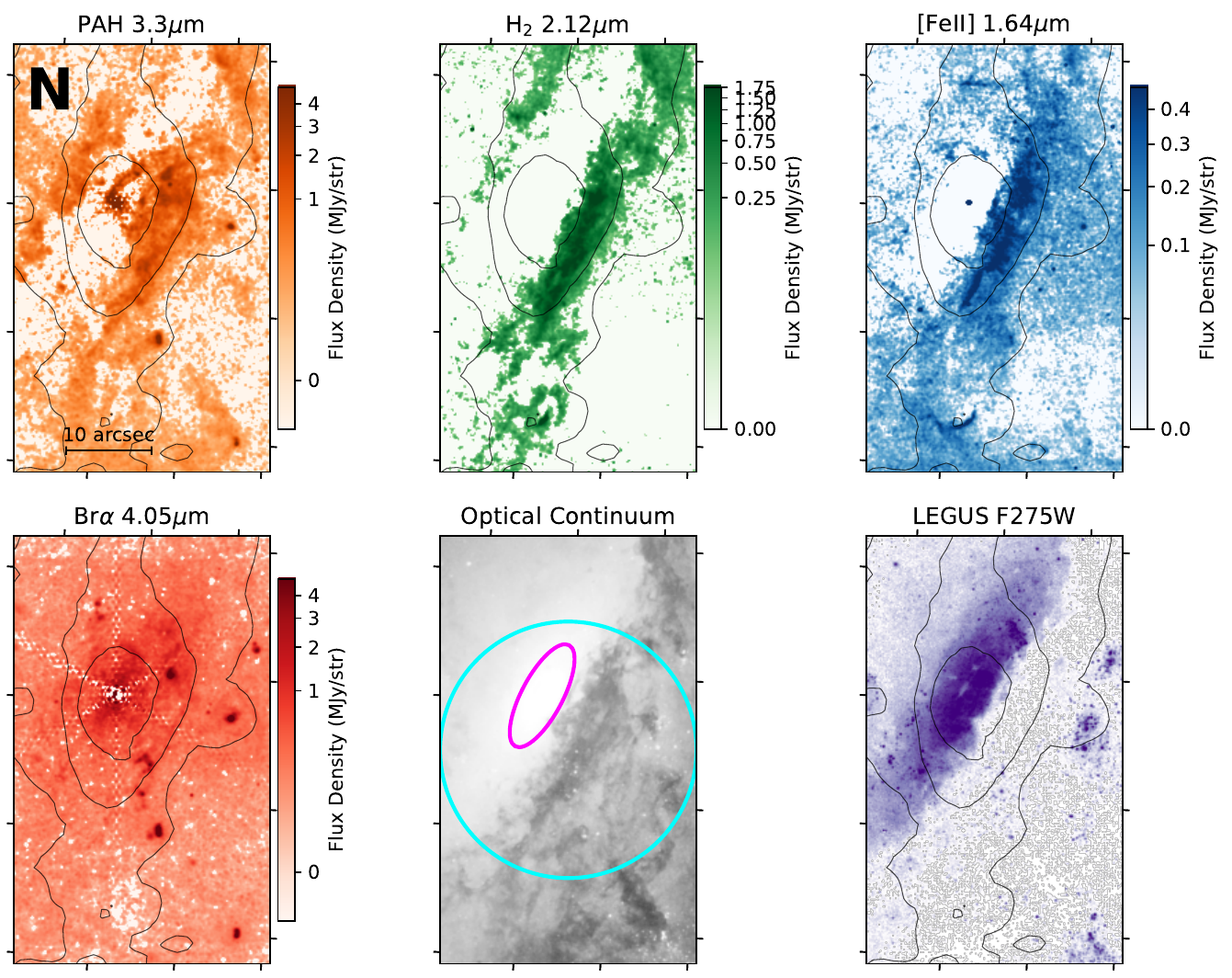}
\caption{
Multi-wavelength comparison of NGC 4258 nuclear region, Region N, observed with JWST and HST. The top row shows smoothed images of the PAH emission at 3.3$\mu$m (left), H$_2$ emission at 2.12$\mu$m (center), and \feii emission at 1.64$\mu$m (right) captured by JWST. The bottom row displays the Br$\alpha$ emission at 4.05$\mu$m (left), optical imaging from HST (center), and HST UV F275W imaging from LEGUS (right). VLA 4.8 GHz emission is overplotted as black contours. The color bars indicate the flux density in MJy/str for each corresponding JWST image, highlighting the spatial distribution of different emission features across the nuclear region. Typical aperture extent used in photometry across each Region is shown in Cyan over the optical imaging, with the masked nuclear region shown in magenta.
}
\label{fig:grid_nucleus}
\end{figure*}

{\it Region N:} Figure \ref{fig:grid_nucleus} shows a $30'' \times 50''$ field centered on the nucleus of NGC 4258. The active nucleus is identifiable by diffraction artifacts in the PAH and Br$\alpha$ maps and appears as an isolated point source in the \feii map. Surrounding the nucleus, there is significant Br$\alpha$ emission. At larger radii, \feii, H$_2$, and PAH emissions trace the morphologies of dust lanes in both the background and foreground relative to the AGN, with the most prominent emitter being a foreground dust lane southwest of the nucleus. Notably, PAH emission directly follows the morphology of this dust lane in several areas. Assuming the AGN is the dominant ionization source, the near-parallel orientation of the dust lanes with our line of sight complicates the detection of stratification between the different emitters in this region. However, small areas of stratification appear directly south of the nucleus. Additionally, an arc or bow shock structure is visible southeast of the nucleus near the bottom of the field, prominent in \feii and H$_2$ emission. This feature coincides with a radio hotspot described by \cite{ogle_jet-shocked_2014}.

\begin{figure*}
\centering
\includegraphics[width=0.98\textwidth]{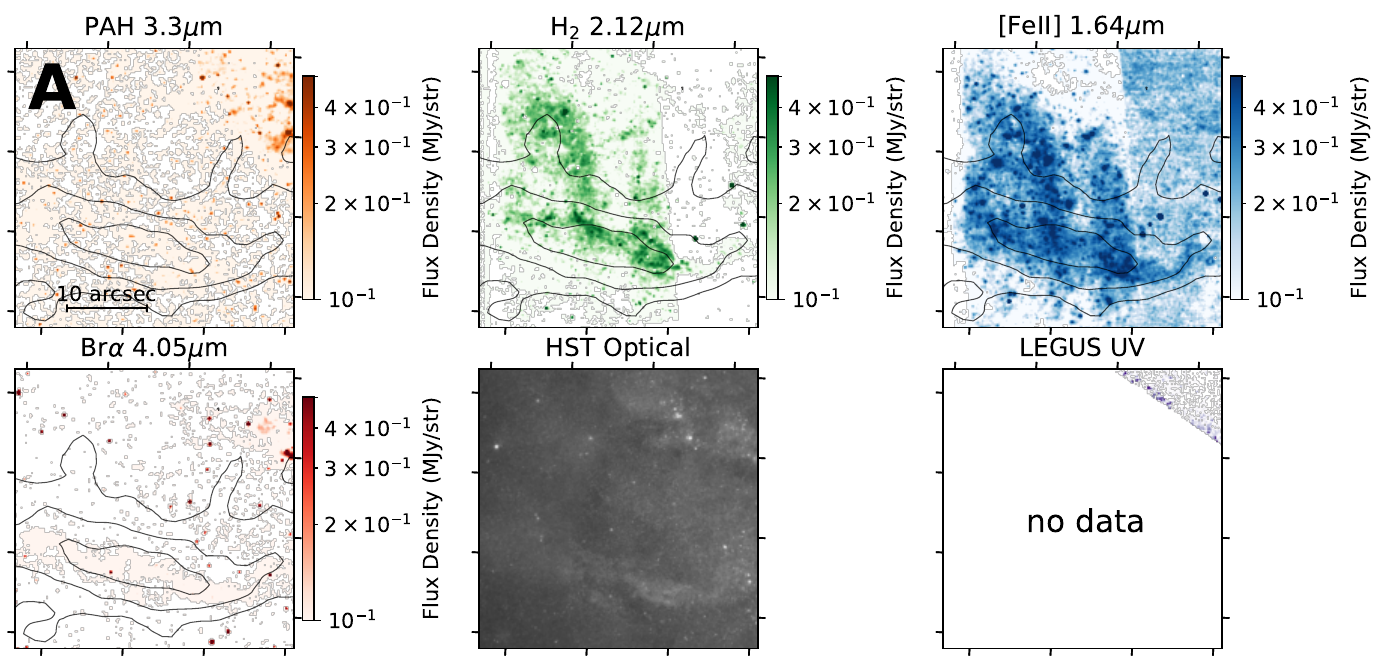}\\
\vspace{1cm}
\includegraphics[width=0.98\textwidth]{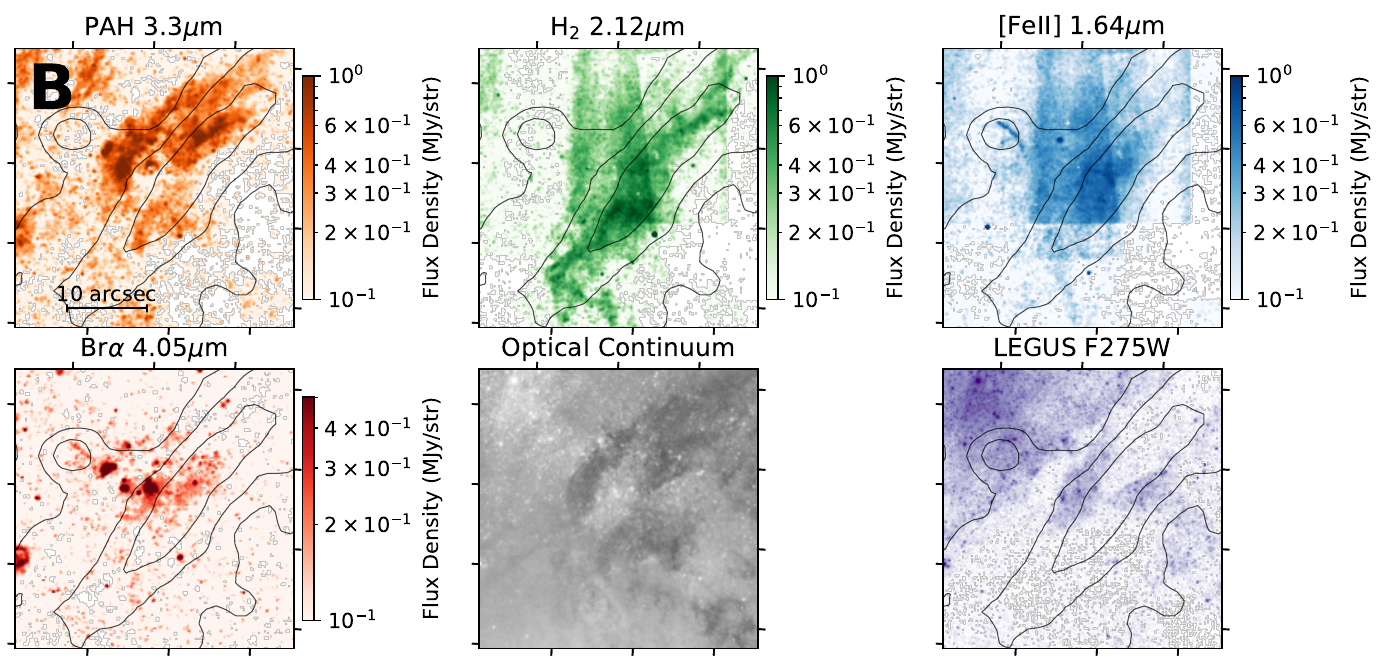}
\caption{
Multi-wavelength comparison of Regions A and B in NGC 4258 observed with JWST and HST. The top row for each Region shows smoothed images of the PAH emission at 3.3$\mu$m (left), H$_2$ emission at 2.12$\mu$m (center), and \feii emission at 1.64$\mu$m (right) captured by JWST. The bottom row for each Region displays the Br$\alpha$ emission at 4.05$\mu$m (left), an optical image from HST (center), and HST UV F275W imaging from LEGUS (right). VLA 4.8 GHz emission is overplotted as black contours. The color bars indicate the flux density in MJy/str for each corresponding JWST image, highlighting the spatial distribution of different emission features across each region.
}
\label{fig:grid_AB}
\end{figure*}

{\it Region A:} Figure \ref{fig:grid_AB} illustrates a $35'' \times 35''$ field encompassing the brightest knot in the southeast anomalous radio arc. PAH emission is predominantly absent in this field, with the exception of a portion of a star-forming region, also visible in Br$\alpha$, in the upper right corner. \feii and H$_2$ emissions are prominent, while Br$\alpha$ is faintly detected along the bottom of the field, coinciding with the location of the bright radio emission as depicted in Figure \ref{fig:ContSubMosaicFeiiH2}, which is also faintly observable in optical imaging. Notably, discrete knots are interspersed among diffuse \feii and H$_2$ emissions to the north of this region. We measure the diameters of several distinct, bright knots, at flux levels exceeding 3$\sigma \times$ rms of the background, to be approximately 0.06$"$ (2.1 pc) - 0.42$"$ (14.9 pc), acknowledging that this does not necessarily reflect their full extent.


{\it Region B:} Figure \ref{fig:grid_AB} also shows a $35'' \times 35''$ field over the brightest knot in the northwest anomalous radio arc. This field is complex due to signatures from both star formation and shocks, in addition to unfortunate imaging artifacts in \feii and H$_2$. PAH emission is colocated with dust lanes visible in the optical, north of the radio flux peak, and absent from dust lanes south of the radio peak. Dust lanes obscure young star formation that is more visible in the north portion of the region from UV imaging. Br$\alpha$ is bright both throughout the PAH emission and also over the radio flux peak. We also see diffuse \feii at the radio peak, with H$_2$ extending northwest and southeast along the radio morphology. The difference in overall \feii morphology here versus what is observed in Region A is notable, with a severe lack of discrete parsec-scale knots. A radio knot to the northeast in Region B marks the location of a second, northern radio hot spot \cite{ogle_jet-shocked_2014}, which shows a sharp and discrete arc in \feii and Br$\alpha$. This arc measures approximately 2.5$"$ (88.7\,pc) long by 0.2$"$ (7.1\,pc) wide in \feii, in which it is brightest. 

\begin{figure*}
\centering
\includegraphics[width=0.98\textwidth]{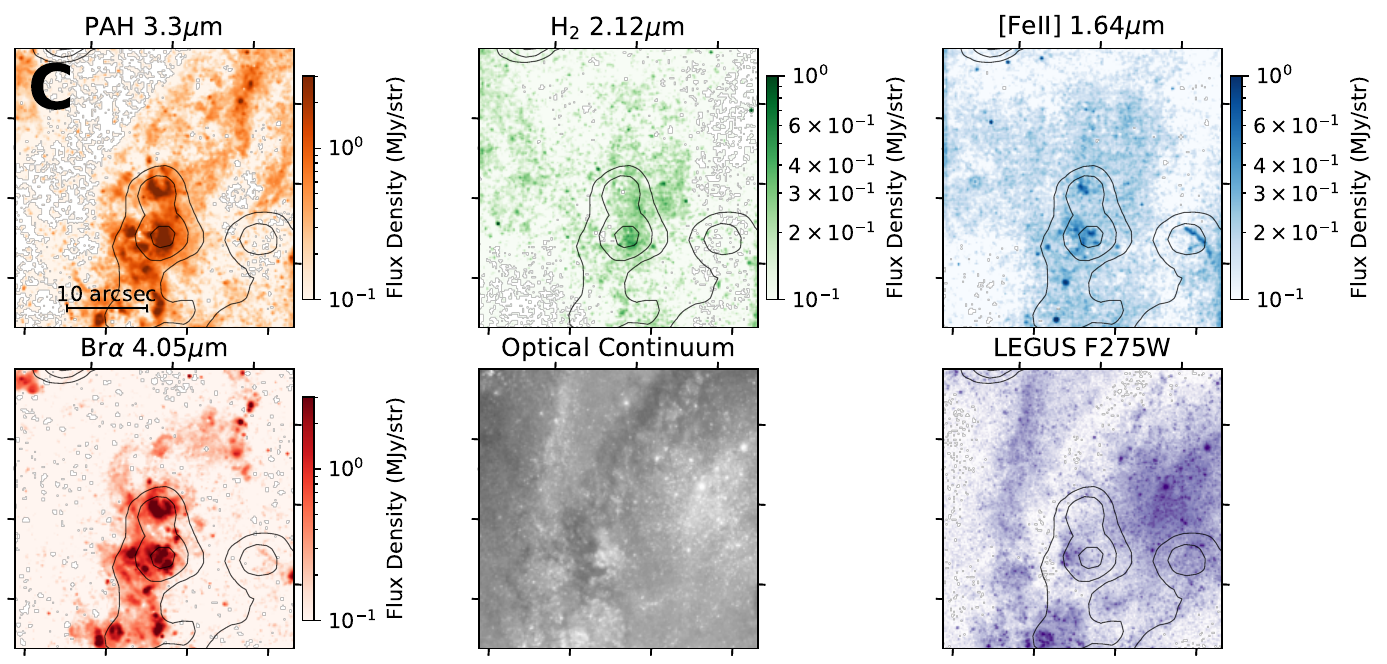}\\
\vspace{1cm}
\includegraphics[width=0.98\textwidth]{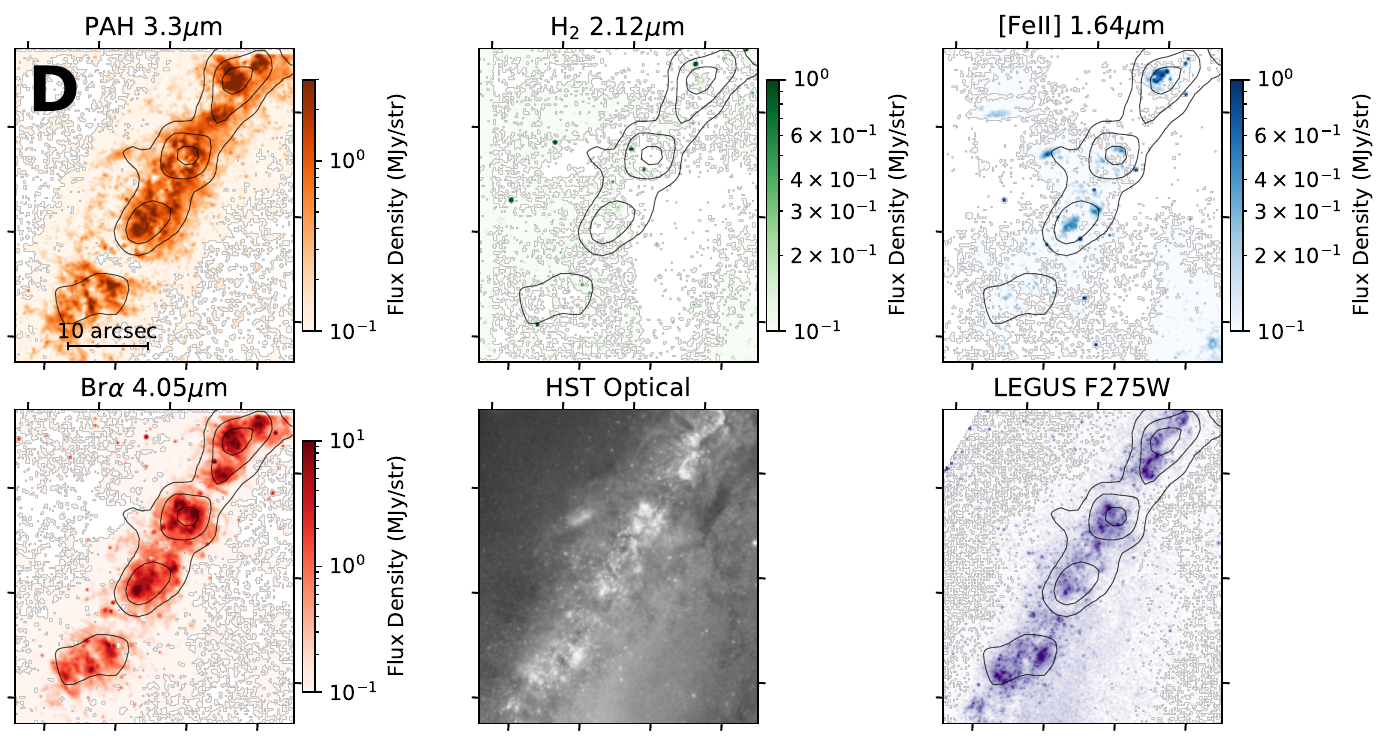}
\caption{
Multi-wavelength comparison of Regions C and D in NGC 4258 observed with JWST and HST. The top row for each Region shows smoothed images of the PAH emission at 3.3$\mu$m (left), H$_2$ emission at 2.12$\mu$m (center), and \feii emission at 1.64$\mu$m (right) captured by JWST. The bottom row for each Region displays the Bracket$\alpha$ emission at 4.05$\mu$m (left), an optical image from HST (center), and HST UV F275W imaging from LEGUS (right). VLA 4.8 GHz emission is overplotted as black contours. The color bars indicate the flux density in MJy/str for each corresponding JWST image, highlighting the spatial distribution of different emission features across each region.
}
\label{fig:grid_CD}
\end{figure*}

\begin{figure*}
\centering
\includegraphics[width=0.98\textwidth]{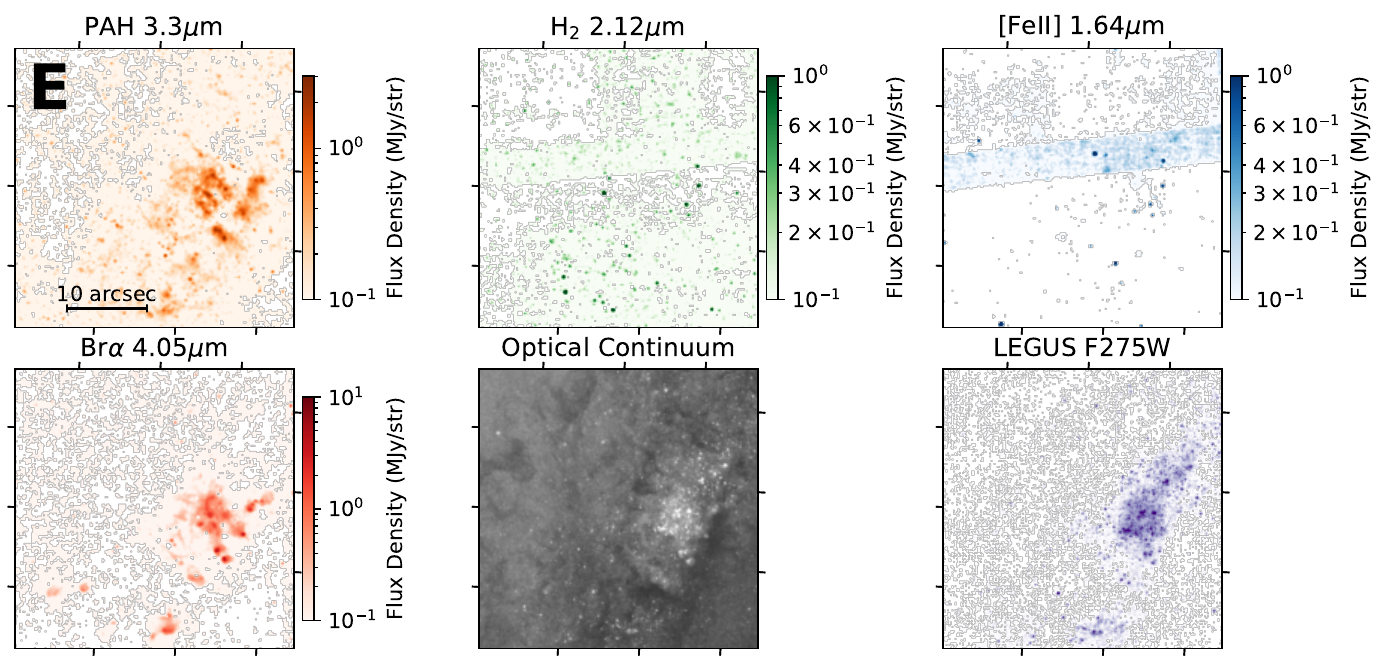}
\caption{
Multi-wavelength comparison of Region E in NGC 4258 observed with JWST and HST. The top row shows smoothed images of the PAH emission at 3.3$\mu$m (left), H$_2$ emission at 2.12$\mu$m (center), and \feii emission at 1.64$\mu$m (right) captured by JWST. The bottom row displays the Br$\alpha$ emission at 4.05$\mu$m (left), an optical image from HST (center), and HST UV F275W imaging from LEGUS (right). All images are aligned to the same celestial coordinates, with the field of view centered at RA = 12h18m57.2s and Dec = +47$^{\circ}$ 18'07.8\arcsec. VLA 4.8 GHz emission is overplotted as black contours. The color bars indicate the flux density in MJy/str for each corresponding image, highlighting the spatial distribution of different emission features across the region.
}
\label{fig:grid_E}
\end{figure*}

{\it Regions C, D, E:} We select three additional regions, shown in Figures \ref{fig:grid_CD}-\ref{fig:grid_E}, that do not significantly overlap with the anomalous radio arc features, but instead highlight star formation occurring in the spiral arms north and south of the nucleus. The findings in each of these fields are consistent: radio knots, when present, overlap exceedingly bright fields of Br$\alpha$ and surrounding PAH emission, which also correspond to bright, resolved UV morphologies indicative of young stellar populations. Comparatively, \feii and H$_2$ fluxes are modest, and distinct flux knots are sparse. As such, the radio emission observed in and around these regions, if at all, represents relatively young supernovae remnants (SNRs) directly associated with star formation \citep{Tur94}. This includes the first radio-detected SNR, designated SNR 1981K \citep{van83b}, which is located just north of Region C, as shown in Figure \ref{fig:fov}.


To measure the flux of emission lines in the JWST images and quantify the differences in each region, we performed aperture photometry using circular apertures centered on the target regions of interest, with nearby background regions used for local sky subtraction. The positions of the target and background apertures were defined in celestial coordinates (RA, Dec) and converted to pixel coordinates using the World Coordinate System (WCS) from the image headers. The photometric apertures had a radius of 15 arcseconds (500 pixels at the JWST pixel scale of 0.03 arcseconds per pixel), chosen to encompass the extended emission regions while minimizing contamination from surrounding structures, and background apertures were 8 arcseconds in radius. An elliptical mask, 6$\farcs$7 $\times$ 2$\farcs$4 with a major axis position angle of 62$^{\circ}$, was applied only over the nucleus (Region N) to address over-subtraction issues, where negative fluxes ("bowling") were prevalent around the nucleus due to steep gradients in emission. For each aperture, we calculated the total flux by summing pixel values and subtracted the background flux, which was scaled to match the aperture area. The resulting net fluxes were converted to Janskys (Jy) using the known pixel scale and solid angle per pixel. Flux uncertainties were determined by propagating the statistical uncertainties in the aperture and background measurements. We provide flux measurements in Table~\ref{tab:table1} and ratios of flux measurements in Table~\ref{tab:line_ratios}.

\begin{table*}
\centering
\begin{tabular}{llcc}
\hline
Region & Coordinates (RA, Dec) [deg] & Emission Feature & Flux [Jy] \\
\hline
\multirow{4}{*}{N} & \multirow{4}{*}{(184.7384983, 47.3021741)} & PAH & $2.96 \times 10^3 \pm 3.33$ \\
 & & $\mathrm{H_{2}}$ & $1.13 \times 10^3 \pm 5.98$ \\
 & & Br$\alpha$ & $3.25 \times 10^3 \pm 2.64$ \\
 & & [Fe II] & $7.55 \times 10^2 \pm 4.10$ \\
\hline
\multirow{4}{*}{A} & \multirow{4}{*}{(184.7729912, 47.2845995)} & PAH & {$2.57 \times 10^2 \pm 0.83$} \\ 
 & & $\mathrm{H_{2}}$ & $1.30 \times 10^3 \pm 4.28$ \\
 & & Br$\alpha$ & $1.87 \times 10^2 \pm 2.84$ \\
 & & [Fe II] & $3.98 \times 10^3 \pm 13.7$ \\
\hline
\multirow{4}{*}{B} & \multirow{4}{*}{(184.7336866, 47.3155555)} & PAH & $2.48 \times 10^3 \pm 1.05$ \\
 & & $\mathrm{H_{2}}$ & $2.74 \times 10^3 \pm 24.4$ \\
 & & Br$\alpha$ & $1.00 \times 10^3 \pm 2.48$ \\
 & & [Fe II] & $3.06 \times 10^3 \pm 4.81$ \\
\hline
\multirow{4}{*}{C} & \multirow{4}{*}{(184.7429631, 47.3195524)} & PAH & $4.09 \times 10^3 \pm 1.05$ \\
 & & $\mathrm{H_{2}}$ & $1.42 \times 10^3 \pm 24.4$ \\
 & & Br$\alpha$ & $3.43 \times 10^3 \pm 2.48$ \\
 & & [Fe II] & $2.49 \times 10^3 \pm 4.81$ \\
\hline
\multirow{4}{*}{D} & \multirow{4}{*}{(184.7347248, 47.3382625)} & PAH & $4.51 \times 10^3 \pm 1.20$ \\
 & & $\mathrm{H_{2}}$ & NA \\
 & & Br$\alpha$ & $5.77 \times 10^3 \pm 18.3$ \\
 & & [Fe II] & $1.10 \times 10^3 \pm 3.45$ \\
\hline
\multirow{4}{*}{E} & \multirow{4}{*}{(184.7425031, 47.2674577)} & PAH & $1.57 \times 10^3 \pm 3.18$ \\
 & & $\mathrm{H_{2}}$ & NA \\
 & & Br$\alpha$ & $7.58 \times 10^2 \pm 4.08$ \\
 & & [Fe II] & NA \\
\hline

\end{tabular}
\caption{Fluxes of PAH 3.3$\mu$m feature, and $\mathrm{H_{2}}$, Pa$\alpha$, and \feii lines in Regions N-E as shown in Figures \ref{fig:grid_nucleus}-\ref{fig:grid_E}. The flux values are expressed as $F \pm \sigma$, where $F$ is the flux and $\sigma$ is the associated uncertainty. 'NA' indicates negligible or undetectable emission for the given line.}
\label{tab:table1}
\end{table*}

\medskip

\begin{table*}
\centering
\begin{tabular}{lcccccc}
\hline
Region & PAH/Br$\alpha$ & PAH/H$_2$ & PAH/[Fe II] & Br$\alpha$/H$_2$ & Br$\alpha$/[Fe II] & H$_2$/[Fe II] \\
\hline
N & 0.91 & 2.62 & 3.92 & 2.88 & 4.30 & 1.50 \\
A & 1.37 & 0.20 & 0.06 & 0.14 & 0.05 & 0.33 \\
B & 2.48 & 0.91 & 0.81 & 0.36 & 0.33 & 0.90 \\
C & 1.19 & 2.88 & 1.64 & 2.42 & 1.38 & 0.57 \\
D & 0.78 & NA & 4.10 & NA & 5.25 & NA \\
E & 2.07 & NA & NA & NA & NA & NA \\
\hline
\end{tabular}
\caption{Ratios of emission feature fluxes for each region. 'NA' indicates that one or more fluxes were negligible or undetectable, preventing calculation of the ratio.}
\label{tab:line_ratios}
\end{table*}

\section{Discussion and Future Work}

Our observations reveal intricate structure and clear stratification within the interstellar medium, particularly along the interface between the observed radio structure and the surrounding gas. Emission from Br$\alpha$, \feii, H$_2$, and PAHs shows distinct stratification across several regions. Notable examples include: \textit{Region N}: A dense foreground dust lane exhibits \feii emission located interior to H$_2$ emission relative to the central AGN. \textit{Region A}: There is a transition from a blend of Br$\alpha$ and \feii emission at the radio peak to a blend of \feii and H$_2$ emission farther north of the peak. \textit{Region B}: \feii emission is concentrated at the peak of the radio structure, with more extended H$_2$ emission surrounding it. In the northern portion of Region B, the extended H$_2$ emission forms a distinct boundary adjacent to bright PAH structures.

\subsection{Radio emission as Mechanical Feedback from Nuclear-Winds}

The progenitors of shocks in a majority of AGN are largely debated as of late. Classically, shocks are attributed to jets, with 'radio-quiet' AGN hosting smaller versions of jets observed in 'radio-loud' AGN \cite{Muk18,Jar19,Ven21,Gir22}. However, recent works argue that shocks can also be produced by AGN-driven winds running into dense host material at greater radii, which in turn produce the observed radio emission as a byproduct \cite{Pan19,Smi20,Fis21,Har24}.

Spatially resolved kinematic data in NGC 4258 are limited \citep{Saw07,appleton_jet-related_2018} but support NGC 4258 being a radiatively weak AGN \citep{Mas22}, as gas reservoirs throughout the inner kpcs, and particularly near the extended radio emission, are largely in rotation. However, even in galaxies hosting relatively weak AGN, we can still expect winds launched at small radii to reach escape velocities (i.e., \citealt{Fis17,Mee23,Fal24}) and, as such, travel outward until they run into a dense medium at greater radii. 

H$_2$ suffers dissociation for velocities in excess of 50 km s$^{-1}$ where $n = 10^2$ cm$^{-3}$ (H, H$_2$ collision partners) and 25 km s$^{-1}$ where $n > 10^4$ cm$^{-3}$ (e$^-$ collision partner) \citep{Hol89}.  It is likely that our observations predominantly probe the $n = 10^2$ cm$^{-3}$ scale, corresponding to giant molecular clouds, which are resolved in NGC 4258, but not dense cloud cores.  The presence of H$_2$ emission puts a loose upper limit on shock velocities.  However, it should be noted that H$_2$ does appear to be robust enough to exist in surprisingly harsh environments \citep{App06,App23}, indicating either a diminishing supply, regeneration, or shielding from the highest shock velocities and excitation by turbulent decay of shock energy input.

\feii is an indicator of shocks with generally higher velocities:  $50$ km$^{-1} \le v \le 300$ km$^{-1}$ (e.g., \citealt{Koo16}).  \feii is often associated with AGN-driven shocks in the literature \citep{Kno96,Oli01,Rif13,Gar24}, although it can also be excited by stellar winds associated with young stars \citep{Rod04, Ott24}.  Additionally, being a refractory element, Fe is typically depleted in the interstellar medium and resides in dust grains. Thus, \feii line emission can be interpreted as evidence of dust destruction. Indeed, several studies have observed an anti-correlation between \feii emission and PAH features in AGN environments, suggesting that shocks not only liberate Fe from dust grains but also contribute to the destruction or suppression of PAH molecules \citep{Gar24}.

The detection of both \feii and H$_2$ emission in several regions across NGC 4258 suggests the presence of multi-phase shocks. Moderate-velocity J-type shocks, with velocities of 100--300 km/s, destroy dust grains and release Fe into the gas phase. These shocks are capable of dissociating molecular hydrogen at the shock front, but H$_2$ can survive or reform in post-shock cooling zones or slower C-type shocks. 

The coexistence of \feii and H$_2$ emission indicates that the progenitor shock likely has a range of velocities, with high-velocity components driving \feii production and lower-velocity components preserving molecular gas. Thus, the velocities of the shock progenitor are likely on the order of one or a few hundred km s$^{-1}$ and not relativistic. This is evidence that there cannot be relativistic motions typical of a radio jet near these shocks, as the molecular hydrogen gas would be destroyed upon impact.

This interpretation is consistent with simulations showing that relativistic AGN jets interacting with dense ISM tend to destroy or evacuate molecular gas through strong shocks and heating \citep{Wag12}. Observations of strong H$_2$ emission in high-velocity, non-relativistic shocks (e.g., Stephan’s Quintet; \citealt{App06, Gui09}) suggest that molecular gas may survive or reform in such environments, but not under the extreme conditions typically associated with relativistic jets.

In contrast, the presence of \feii emission alongside Brackett-$\alpha$ emission suggests a more energetic and destructive shock scenario. Brackett-$\alpha$ emission is typically associated with photoionized gas in H~\textsc{ii} regions, where ionizing radiation from young stars or an AGN ionizes hydrogen atoms \citep{Dop05}. However, it can also arise in fast J-type shocks (velocities $\gtrsim$ 300 km s$^{-1}$) where collisional ionization and heating are sufficient to produce hydrogen recombination lines alongside forbidden line emission \citep{All08}. In such environments, shock velocities may be too high to preserve molecular hydrogen, resulting in the absence of H$_2$ emission.

In our observations, we find that star-forming regions, identified by strong UV emission and coincident PAH and Br$\alpha$ emission, are clearly distinguishable from regions that exhibit Br$\alpha$ and \feii emission but lack UV point sources and have little or no PAH emission. These latter regions likely trace shocked gas rather than stellar photoionization, as the absence of young stars and the suppression of PAH emission both point toward a different excitation mechanism. This observational distinction helps disentangle the origin of the recombination line emission, which is otherwise ambiguous without velocity diagnostics \citep[e.g.,][]{Dav07}.

Similar approaches have been used in previous spatially resolved studies of nearby AGN, where combinations of recombination, forbidden, and molecular lines are used to separate shock excitation from star formation, including in galaxies like NGC 1068 and NGC 4151 \citep[e.g.,][]{storchi-bergmann_feeding_2009, Dag19}.


\begin{figure*}
\centering
\includegraphics[width=0.98\textwidth]{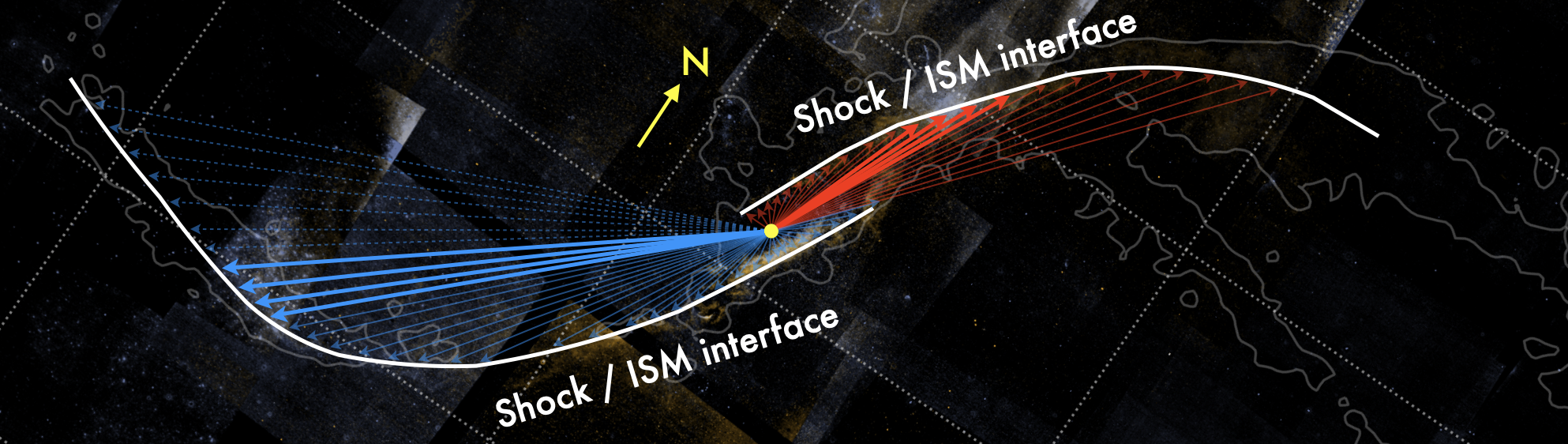}
\caption{Schematic illustration of the proposed outflow-driven shock scenario in NGC 4258. Blue and red arrows represent the path of outflows from opposite sides of the AGN. Thick arrows indicate regions where \feii emission is observed, tracing the most energetically shocked areas. These \feii regions appear roughly 180$^{\circ}$ apart and are encased by lower-energy H$_2$-emitting regions (indicated by thin solid arrows). Dashed blue lines represent areas where H$_2$ observations are unavailable. The curved white boundary marks the shock/ISM interface, showing how the outflows interact with the surrounding interstellar medium.
}
\label{fig:cartoon}
\end{figure*}

Assuming a bipolar outflow typical of radiative AGN feedback, the winds emanate from angles along our line of sight near the nucleus and extend past the peaks of the \feii flux and the extended radio emission in a clockwise pattern, as illustrated in Figure \ref{fig:cartoon}. The most energetically shocked regions, identified by the \feii emission, are found roughly 180° apart from each other. These areas are surrounded by lower-energy H$_2$ regions, all of which can be traced back to the nucleus in Region N. Therefore, we propose that the H$_2$ and \feii emissions align with a scenario in which relatively low-velocity outflows create the observed extended radio structure.

Although models of radio jets have been used to explain the observed radio structure in radio-quiet AGN such as this one (e.g., \citealt{Muk18}), the actual morphology is likely influenced by interactions with host material. In this case, the nuclear emission primarily originates from the nearby dust lanes rather than the central engine. Very Large Array (VLA) A-array observations at 5–22 GHz reveal that the nuclear radio emission is compact and unresolved on scales smaller than $\sim$15 pc, with no evidence of extended collimated structure \citep{Doi13}. For comparison, the jet in M87, one of the best-studied radio-loud AGN, has a transverse width of $\sim$20–100 pc within the inner kiloparsec \citep{Asa12}. At the resolution afforded by A-array for NGC 4258, a similarly narrow, collimated jet would be readily detectable. Its absence, therefore, suggests that the extended radio morphology seen at lower resolution arises not from a classical jet, but from broader interactions between the AGN and host ISM.


The lack of a collimated jet at the nucleus in high-resolution 4.8 GHz maps, combined with the disturbed morphology of the radio arms, indicates that the radio emission is produced by an AGN wind impacting the disk rather than originating from a single precessing jet. In other words, the irregular arms likely represent the points where the AGN outflow strikes dense gas in the disk, leading to shocks and the creation of synchrotron-emitting plasma. This interpretation eliminates the need for dramatic jet precession to explain the observations; instead, the geometry and clumpy structure of the disk shape the appearance of the radio arms.

This perspective aligns with recent studies of winds in radio-quiet AGN \citep{Smi20, Fis23, Che24, Har24}, which show that these winds can inject turbulence and drive shocks without forming a well-collimated relativistic jet. The curved and asymmetric nature of the arms in NGC 4258 fits into this framework, suggesting that the AGN-driven outflow interacts with dense clouds in the interstellar medium (ISM), disrupting its trajectory.

Similar processes are likely responsible for the extended radio emission observed in other nearby radio-quiet AGN. For example, the prototypical Seyfert 2, NGC 1068, has an extended, lobe-like radio structure that was suggested by \cite{Fis23} to result from winds launched from small radii, excavating an adjacent dust lane and producing the observed synchrotron emission as a shock byproduct. NGC 7319, an AGN in Stephen's Quintet observed with JWST \citep{Per22}, has radio emission concentrated near the nucleus and then at two extended regions of different radii, similar to NGC 4258, where evidence suggests that corresponding \feii is strongest. In general, spatially resolved analyses of \feii and radio continuum \citep{storchi-bergmann_feeding_2009,riffel_feeding_2014,Bar14,Fis23,Das24} continue to provide similar findings: these signatures are enhanced where AGN feedback interacts with the intrinsic ISM and depends on the orientation between the AGN and the said ISM.

Recent modeling of H$_2$ emission from shocked gas \cite{Kri23} shows that dissociative J-type shocks produce a spectrum dominated by \textit{vibrational} H$_2$ lines. In contrast, non-dissociative C-type shocks produce a spectrum dominated by \textit{rotational} H$_2$ lines. To differentiate between the two shock types and quantify the amount of energy being injected into the ISM \cite{Neu06,Mar09}, the full spectrum of both rotational and vibrational H$_2$ lines is required. This necessitates both MIRI and NIRSpec observations to probe the rotational and vibrational transitions of H$_2$, respectively. The \feii lines measured by both NIRSpec and MIRI will trace a range of shock velocities and densities, offering further insights into the ionization state and temperature of the shocked gas \cite{Koo16} and providing key information about the physics of the shocks.

Further data are needed to confirm the nature of the shocks. NIRSpec and MIRI IFU spectroscopy would complement the imaging by capturing the full suite of shock diagnostics and quantifying the energy reprocessed in molecular gas across the excitation range of H$_2$. As shown in \cite{Kri23}, flux ratios between H$_2$ lines offer a clear diagnostic for determining shock origin, whether mechanical or radiative. With both near- and mid-IR spectra, one could trace how AGN-driven shocks dissipate energy into the interstellar medium, providing critical insights into feedback mechanisms in NGC 4258.

\begin{figure*}
    \centering
    \includegraphics[width=0.9\textwidth]{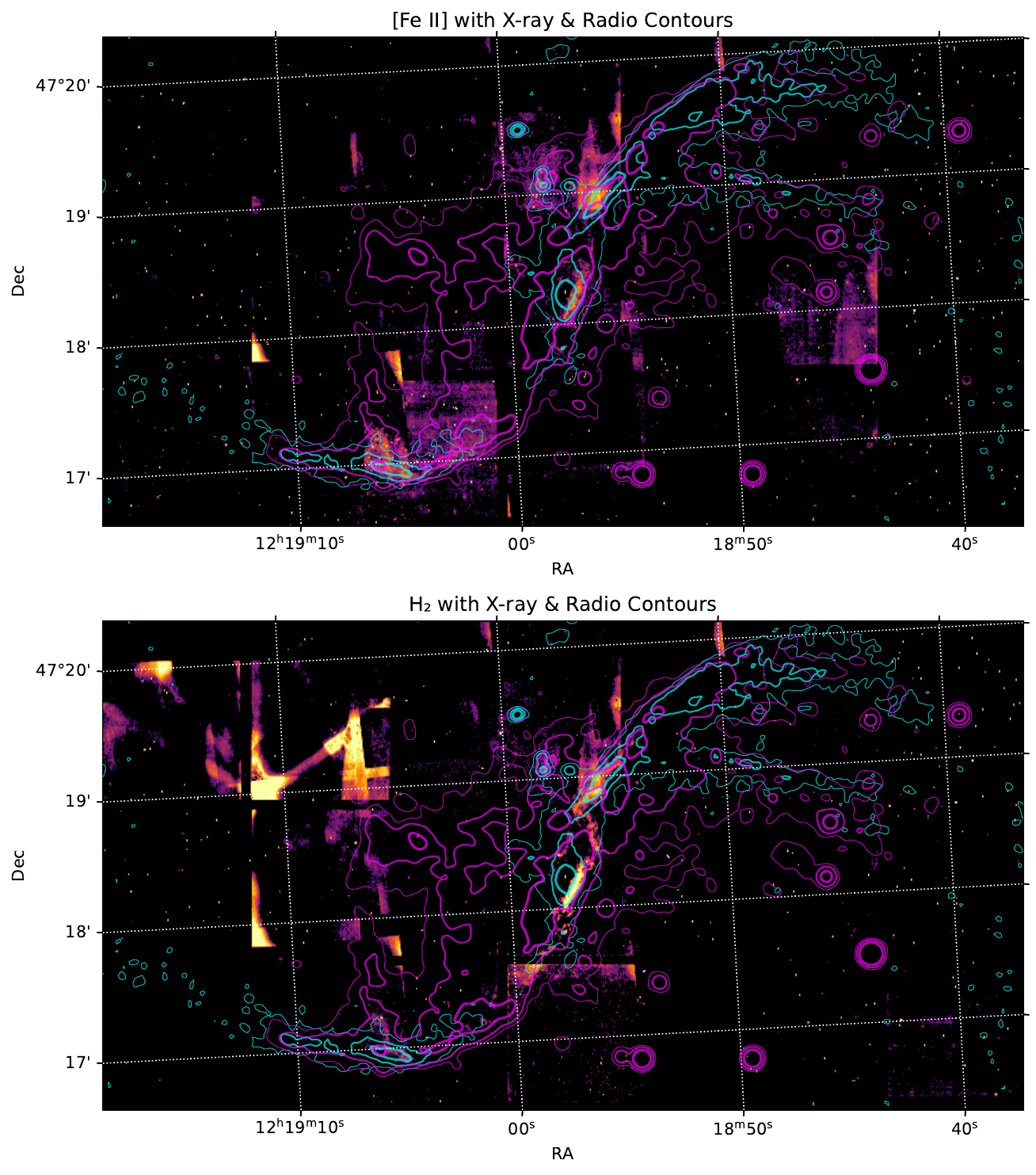}
    \caption{Comparison of [Fe II] and H$_2$ emission from JWST imaging with Chandra X-ray and VLA radio emission. The \textit{top panel} shows [Fe II] emission, while the \textit{bottom panel} presents H$_2$ emission, both mapped onto the same field of view. The magenta contours represent X-ray emission, highlighting regions of high-energy activity, while the cyan contours trace radio emission. The varying contour thickness reflects emission intensity, with thicker lines indicating higher flux levels. JWST imaging artifacts are visible in both images, particularly the striping patterns and abrupt intensity variations. These features are consistent across both datasets, suggesting systematic origins rather than astrophysical structures.}
    
    \label{fig:xray}
\end{figure*}

\subsection{Suggestive Evidence for PAH Destruction by Nuclear-Wind Generated Shocks}

Our observations aimed to determine whether PAH molecules are destroyed by shocks where AGN feedback, traced by the radio emission, impacts the interstellar medium. Evidence for shock dissociation of PAHs has been observed in the Milky Way \citep{Mic10} and JWST presents an opportunity to observe it in other, nearby galaxies (i.e., PHANGS survey of nearby star-forming galaxies: \cite{Cha23, San23}). 

PAH molecules are likely to be destroyed in high-velocity shocks.  For example, \cite{Mic10} show that PAHs with $\sim50$ carbon atoms do not survive in shocks with velocities greater than 100 km s$^{-1}$.  Larger PAHs, those with $\sim200$ carbon atoms, are destroyed in shocks with velocities greater than 125 km s$^{-1}$. In the intermediate $75 - 100$ km s$^{-1}$ range, destruction is not complete but the PAH molecular structure is likely to be denatured by the loss of a significant fraction of atoms. Thus, we expect little PAH emission where AGN-driven wind impacts the interstellar medium unless shielding is present (although they can survive in environments where local dust or gas structure provides shielding from direct mechanical interaction). For example, \citealt{Alo14} detected nuclear 11.3\,$\mu$m PAH emission in several local AGN, arguing that compact, dusty structures such as nuclear gas disks or the torus may shield PAHs from the AGN radiation field while still permitting excitation by other UV sources. These observations underscore the importance of spatial configuration and anisotropic radiation fields in determining PAH survival near AGN.

We compare regions of the interstellar medium excited by the nuclear wind to those associated with star formation, which are typically characterized by slower and more spatially confined outflows. In the former, we expect velocities exceeding 100 km s$^{-1}$. We identify shocks using H$_2$ and \feii emission and examine the presence or absence of associated PAH $3.3~\mu$m band emission.


Our multiband NGC 4258 JWST observations enable us to compare PAH emission in AGN-wind and stellar-wind driven shocks.  We focus on the lettered regions identified in Figure \ref{fig:fov}. Region N is the nuclear region and Regions A and B have concentrations of radio emission and are where it appears that the nuclear wind is being deflected (possibly out of the plane of the galaxy) by impacting dense interstellar medium -- giant molecular clouds.  Regions C and D have concentrations of radio emission but are also sites of intense star formation as indicated by the PAH and Br$\alpha$ emission in Figure \ref{fig:ContSubMosaicPAHBr} and we attribute this to non-thermal emission from young stars \citep{Lau22}. In these locations, the radio likely arises from star formation and not the nuclear wind.  Regions E and F also contain giant molecular clouds and star formation, but weak or absent radio emission compared to the other regions we identify for this analysis.  Because they have weak or absent H$_2$ and \feii emission, we exclude them from this analysis.

Region N:  The nuclear region (Figure \ref{fig:grid_nucleus}) was described in the previous subsection, and the geometry, or more precisely, our viewing angle, makes it difficult to infer the effect of the nuclear wind on PAH emission.  The southwestern dust lane is in the foreground of the nucleus and PAH emission, H$_2$ emission, and \feii emission are all copious, although they likely do not occupy the same physical volumes.  PAH emission is also present in the northeastern dust lane, which is illuminated by the nuclear stars, although the H$_2$ and \feii are relatively weaker.  We consider this region inconclusive.

In Regions A and B, a nearly east-west shock front is noticeable in H$_2$, \feii, and likely Br$\alpha$. This aligns with a hotspot in radio emissions. Additionally, we observe PAH emission in the northwestern corner of the image, while dust lanes appear in the HST optical image, both in the area where PAH emission occurs and in the southern third of the image. Notably, PAH emission is absent at the shock front, which may suggest either PAH destruction due to the shock or a lack of dust in that region. However, the extinction seen in the HST optical image indicates that dust is indeed present. It is also possible that PAHs exist in that area but are not illuminated, preventing them from fluorescing. Overall, we propose that these observations provide compelling evidence for the destruction of PAHs by shocks. However, we note that Region B is one of the most artifact-affected areas in the dataset, with visible striping and low-level continuum subtraction residuals, particularly in the \feii and H$_2$ maps. While these effects complicate quantitative analysis, the broad spatial morphology of the east-west shock front remains clearly detected across multiple tracers and does not spatially align with the processing artifacts. We therefore interpret the observed emission structures as physically meaningful, but caution that detailed photometry in this region may be subject to greater uncertainty than in cleaner parts of the mosaic.

Region B is another area where the nuclear wind is likely interacting with the interstellar medium, as indicated by the enhanced radio emissions, bright H$_2$ emissions, and \feii emissions. There is also PAH emission and dark lanes caused by dust extinction. With the wind flowing from the south and bending northwest, the observed morphology aligns with a shock front corresponding to the H$_2$ and \feii emissions. 

Northeast of the shock near Region C, PAH emission is bright, whereas south of the shock, it is diminished. This could suggest evidence of PAH destruction. It is also possible that there is no source of UV or optical illumination in the south, or that the nuclear wind is not affecting the interstellar medium in that region. However, the latter scenario is unlikely since the wind is not highly collimated. We propose that this could further indicate PAH destruction. 

Regions C and D:  These regions have star formation, indicated by the extended PAH and Br$\alpha$ emission.  H$_2$ and \feii emission are present also, but because these regions are not obviously in the path of the nuclear wind, it is likely that the shocked gas results from stellar winds associated with the star formation.  Where shocked gas is present, it appears to be broadly associated with the PAH emission, in support of a stellar-wind excitation hypothesis.  These regions form a ``control sample" in which shocked gas is present -- from stars, presumably with lower velocities than the nuclear wind -- and PAHs are not generally destroyed.  

Together, these observations suggest that PAH destruction occurs in some regions affected by nuclear-wind-driven shocks, but not in those influenced by stellar-wind-driven shocks. By incorporating HST UV imaging from the LEGUS survey \citep{Cal15}, we find that PAH molecules fluoresce only in dust lanes adjacent to UV-emitting star-forming regions or the AGN. In contrast, several dust lanes—such as those west of the nucleus in the foreground and east of the nucleus in the background—lack nearby UV emitters and do not exhibit PAH fluorescence. This pattern suggests that PAHs emitting at 3.3 $\mu$m, while likely widespread in dust lanes throughout the galaxy, are only excited in regions exposed to UV radiation, similar to snow appearing to be illuminated only over a string of holiday lights. Similar conclusions have been reached in PHANGS-JWST studies of nearby star-forming galaxies, where 3.3\,$\mu$m PAH emission is strongly correlated with UV-bright star-forming regions and largely absent in dust lanes lacking local UV sources \citep[e.g.,][]{Rod23, San23, Sut24, Ujj24}.

In Regions A and B, and to some extent in Region N, we likely observe a combination of X-ray-dominated regions (XDRs) or photon-dominated regions (PDRs) \citep[e.g.,][]{Wol22} along with shocks (Figure \ref{fig:xray}. Both H$_2$ and \feii emission are collisionally excited, originating from winds traveling outward from the AGN and impacting the surrounding medium. However, the excitation conditions differ between these tracers.

H$_2$ emission is more spatially extended because it is excited at lower shock velocities ($\sim$10–50 km s$^{-1}$), allowing molecular hydrogen to survive and re-radiate energy \citep{Hol89, All08}. In contrast, \feii emission is more compact because its excitation typically results from faster shocks ($\gtrsim$50 km s$^{-1}$) that liberate iron from dust grains and produce ionized post-shock gas, which can also dissociate H$_2$ in the same regions. Observations of nearby AGN such as NGC 6240 and NGC 1068 reveal this spatial segregation: H$_2$ traces extended molecular gas, while \feii is concentrated in compact, high-excitation shock zones \citep{van93, storchi-bergmann_feeding_2009}. Consequently, \feii emission is strongest near the brightest regions of the radio morphology, marking the sites of intense, localized shock activity. Meanwhile, the more extended H$_2$ emission traces lower-energy shocks or oblique interactions with the surrounding ISM. This stratification supports a scenario where AGN-driven winds interact with a multiphase ISM, generating shocks with a range of velocities that sculpt the excitation structure of the galaxy.

Although these findings remain suggestive, they demonstrate that JWST’s angular resolution is sufficient to directly observe PAH destruction in a galaxy located 7 Mpc away—an analysis that was previously not possible with {\it Spitzer} \citep{ogle_jet-shocked_2014}.

\subsection{Potential Parallels to Milky Way Fermi Bubbles}
The extended radio structure in NGC 4258 has historically been interpreted as evidence of a weak AGN jet \citep[e.g.,][]{Cec00}. However, our observations suggest that the radio features are largely a result of shocks generated as AGN-driven winds interact with the ambient ISM versus relativistic plasma outflows. This interpretation is potentially analogous to the formation of the Fermi Bubbles in the Milky Way, where a past episode of nuclear activity inflated large-scale lobes visible in radio and X-rays due to shock heating and interactions with the surrounding medium \citep{Su10}. 

The lack of a collimated radio jet at high resolution, combined with the disturbed morphology of the radio arms, supports a scenario where the AGN wind impacts dense gas in the disk, driving shocks that create synchrotron-emitting plasma. The observed spatial correlation between \feii and H$_2$ emission with radio features indicates that these regions are sites of active shock interactions rather than direct jet ejection. This scenario is consistent with models where low-luminosity AGN (LLAGN) operate in an advection-dominated accretion flow (ADAF) mode, producing thermally driven outflows rather than highly relativistic jets \citep{Yua14}.

Radio polarization position angles and spectral index measurements could provide further evidence to distinguish between a thermally driven outflow and a relativistic jet. VLA observations could be used to directly test whether the radio emission originates from a thermally driven wind or relativistic plasma ejection. Such an analysis is beyond the scope of this study and will be addressed in future work.

In this framework, the observed radio structures represent regions where AGN winds have encountered dense gas, compressing and heating the ISM. The resulting shocks excite Fe and molecular hydrogen, leading to the strong \feii and H$_2$ emission we observe. This is distinct from classical radio-loud AGN, where relativistic jets directly produce synchrotron emission. Instead, the radio structures in NGC 4258 appear to be a byproduct of mechanical energy deposited into the ISM by AGN winds, much like the Milky Way's Fermi Bubbles.

This interpretation suggests that LLAGN contribute to feedback primarily through shock-driven turbulence and energy injection rather than through powerful jets. While AGN X-ray and UV radiation also deposit energy into the ISM, their influence is typically more diffuse and sensitive to local gas column densities, as the penetration of ionizing photons depends strongly on the opacity of intervening material \citep{Hop12,Dra11}. In contrast, mechanical energy from outflows can directly compress and shock dense gas, exciting emission lines such as \feii and H$_2$ in the impacted regions \citep{Wag12, Ric18}.

If this scenario is correct, it reinforces the idea that AGN feedback in many Seyfert galaxies and LLAGN may be fundamentally different from the jet-dominated paradigm often associated with AGN-driven outflows. Similar processes may be at work in other AGN, including NGC 1068 and NGC 7319, where \feii and radio continuum correlations indicate a wind-driven, rather than jet-driven, feedback mechanism \citep[e.g.,][]{Bar14,Per22}.

Future JWST NIRSpec and MIRI IFU spectroscopy will allow for a detailed examination of the excitation mechanisms in these regions, distinguishing between mechanical shock and radiative contributions and further testing whether the observed radio structure is indeed a consequence of AGN wind interactions with the ISM.

\subsection{Implications of Mechanical Feedback in Surveys of AGN/Host Coupling}

\begin{figure*}
\centering
\includegraphics[width=0.98\textwidth]{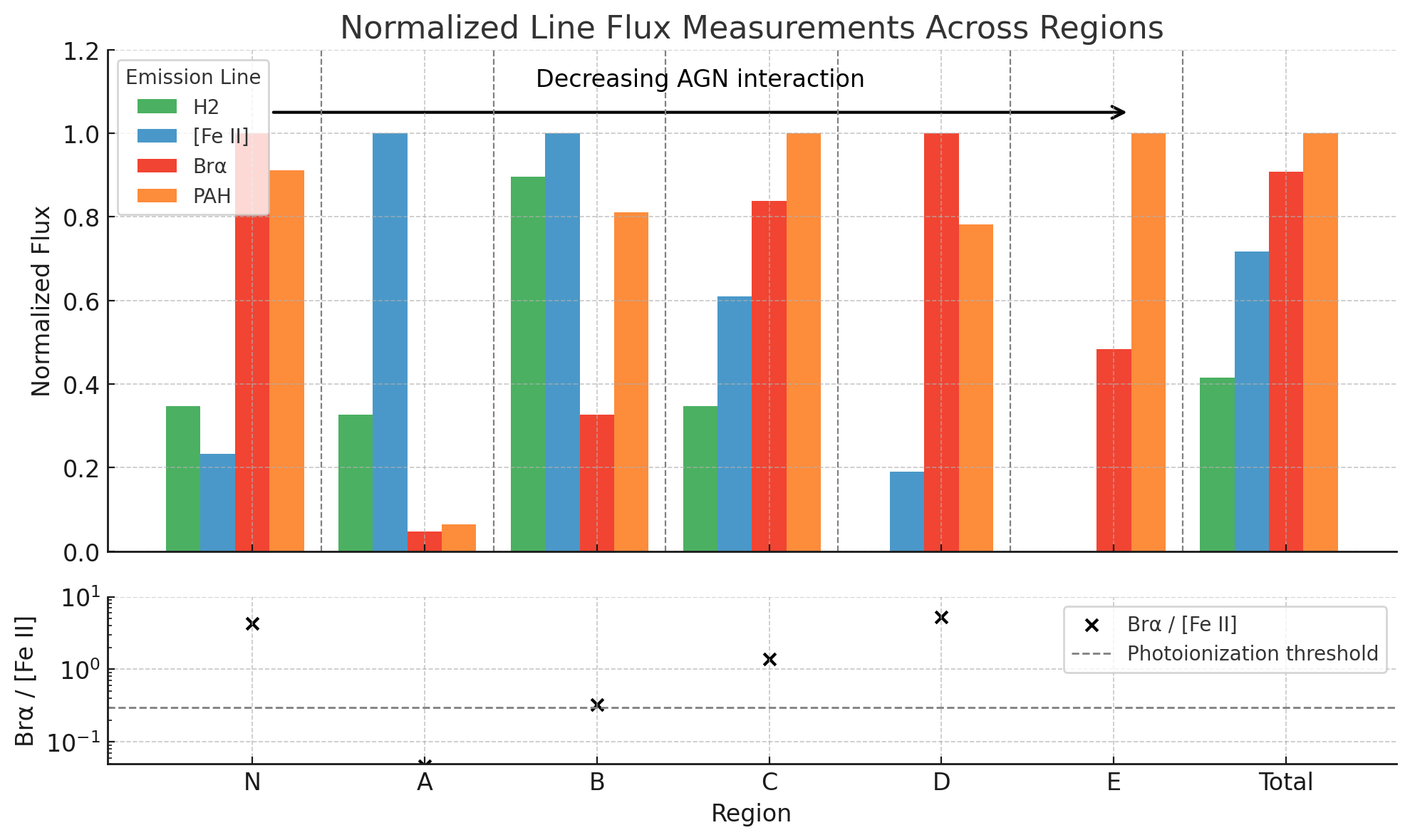}
\caption{Top panel: Bar chart showing normalized emission feature flux measurements from Table 2 across the six measured regions in Figures 6–9 for the four imaged emission features: PAH (orange), Br$\alpha$ (red), H$_2$ (green), and \feii (blue). Fluxes are normalized within each region to the brightest detected feature, and an additional ``Total" column on the right shows relative line strengths integrated across all six regions. Regions with missing flux measurements indicate negligible or undetectable emission for the corresponding feature.  Bottom panel: Scatter plot of the Br$\alpha$/\feii ratio for each region (N through E), plotted on a logarithmic scale. The dashed horizontal line at 0.3 indicates an approximate threshold above which photoionization is expected to dominate over shock excitation \citep{Alo97, Ost06}.
}
\label{fig:barchart}
\end{figure*}

AGN-driven outflows, prevalent across various gas phases (neutral, molecular, and ionized), are thought to be linked to the AGN luminosity and may have different properties depending on these phases (e.g., \citealt{Fio17}). Despite this, the overall effect of AGN feedback on host galaxies remains unclear, necessitating a detailed, quantitative assessment of outflow properties in all phases to constrain theoretical models of feedback. A critical factor in understanding AGN feedback is determining whether there is a favorable geometrical alignment between AGN outflows and surrounding host material (e.g., the host disk or in-spiraling lanes of gas and dust). This alignment, often referred to as “strong coupling,” is crucial for the interaction of AGN feedback with the ISM of the galaxy (see \citealt{Ram22}). Only in cases where AGN feedback is aligned with the ISM do we expect significant coupling and observable interactions.

Local and intermediate-redshift AGN provide excellent astrophysical laboratories for studying the connection between the AGN nucleus and its host galaxy, thanks to the ability to achieve spatial resolution on relevant scales. However, in the dusty circumnuclear regions of AGN, optical observations are often severely obscured by dust, making the infrared spectrum particularly valuable for probing these inner regions. Recent JWST observations of nearby Seyfert galaxies, for instance, suggest that key properties of outflows, such as mass outflow rates, may be underestimated in optical studies due to extinction effects \citep{Her24}. Furthermore, these infrared observations enable deeper investigations into the effects of AGN feedback on the surrounding gas and dust within the host galaxy \citep[e.g.,][]{Lai22,Gar22,Gar24,Dav24}.

In our study, we find that \feii emission arises from the interaction of low-velocity AGN winds with the host interstellar medium at projected distances well within the AGN radiation field, suggesting that the AGN is mechanically depositing energy into its surroundings. Models predict that Br$\alpha$/\feii ratios are elevated in photoionized star-forming regions and suppressed in fast shock-dominated environments, with typical thresholds of $\gtrsim 0.3$ for photoionization and $\lesssim 0.2$ for shocks \citep{Alo97, Ost06}.

We present normalized aperture photometry in Figure \ref{fig:barchart}, showing that \feii is strongest—outside the nucleus—in the shocked Regions A and B, which also coincide with enhanced radio emission and molecular hydrogen features. In contrast, Regions C, D, and E exhibit progressively lower \feii fluxes, which we interpret as reduced direct impact from the AGN wind. Across all apertures, \feii remains relatively strong compared to Br$\alpha$, consistent with widespread shock excitation rather than pure photoionization.

These results suggest that a spectrum of this object at higher redshift, where a single aperture may encompass a larger portion of the galaxy, could yield a positive detection of relatively strong \feii emission. Consequently, while \feii has already been identified as a useful diagnostic for identifying AGN in spectral surveys \citep{Alo97}, our findings indicate that \feii strength could also serve as a valuable indicator of strong coupling between AGN feedback and the ISM, particularly in future surveys of active galaxies.

\subsection{Shock morphology at high resolution}
\label{sec:highres}



One of the most striking features in these observations is the granular morphology of the \feii and H$_2$ emission in Region A. Unlike the diffuse ionized gas seen in AGN photoionization, such as [O~III] emission in other systems \citep{Sch03,Fis18,Sto18,Ven21,Pol24}, the \feii emission in NGC 4258 is resolved into parsec-scale knots, suggesting an origin in shock-excited gas. This substructure is not typically observed in IFU studies of other AGN, where lower spatial resolution limits the ability to distinguish individual shock sites \citep{storchi-bergmann_feeding_2009,Bar14,Per22,Das24}.


The granular \feii morphology in Region A likely arises from variations in shock strength across the dust lane, as the AGN wind collides with dense ISM clumps. At the radio flux peak, \feii emission is strong, and Br$\alpha$ is present at lower levels, suggesting either high-energy shocks or photoionization contributing to the excitation. Moving northward from the radio peak, Br$\alpha$ diminishes, while H$_2$ emission strengthens, consistent with lower-energy shocks preserving molecular gas.

A similar stratification is tentatively seen in Region B, although imaging artifacts make it difficult to determine whether the \feii emission is as granular as in Region A. The differing morphologies in Regions A and B may reflect differences in the shock propagation angle. In Region A, the AGN wind appears to impact the dust lane at a nearly perpendicular angle, leading to compact, localized shocks. In Region B, the interaction may be more oblique, distributing the shock energy over a broader area and producing a smoother \feii distribution.

These observations highlight how spatially resolved feedback studies in nearby AGN can inform our understanding of unresolved AGN winds at higher redshifts. Further spectroscopic analysis is needed to compare the kinematics of these shocked regions with the disturbed, high-velocity gas observed in lower-resolution IFU studies \citep{Per22,Das24}. The ability to resolve individual shock structures in NGC 4258 provides a valuable comparison to more distant systems, where spatially unresolved outflows may blend together, affecting nuclear measurements and interpretations of feedback strength.

\subsection{Future Work}  
\label{sec:futurework}  

This paper is the first in a series of studies on NGC 4258, providing a broad overview of this rich JWST dataset. Several follow-up analyses are planned to explore specific aspects of the galaxy in detail. One of the key next steps is the investigation of H$_2$ and CO absorption to trace the excitation conditions of molecular gas and study the role of molecular clouds in the central regions of NGC 4258. Building on our continuum-subtracted emission maps, we will generate extinction and reddening maps to better understand dust distributions and their relationship to the observed emission features. In parallel, we will perform a detailed analysis of stellar populations, focusing on young stellar objects (YSOs) and their spatial distribution relative to shock-excited regions.

A critical part of this future work will be the in-depth analysis of shock models, with a focus on deriving shock velocities responsible for the observed \feii and H$_2$ emission. Ratios of \feii to hydrogen recombination lines, such as Pa-$\beta$, serve as powerful diagnostics of AGN-driven photoionizing shocks. Following the methodology of \cite{Alo97}, we will investigate these ratios to differentiate between shock excitation in regions dominated by AGN, starbursts, and supernova remnants. This will include mapping the radial distributions of these emission lines while accounting for projection effects within the plane of the host galaxy.

We will also explore how high-resolution JWST, HST, and VLA images of NGC 4258 can serve as analogs for more distant galaxies, helping to interpret observations of similar systems at higher redshifts. Future work will incorporate gas kinematics using long-slit and integral field unit (IFU) spectroscopy to reveal how the gas dynamics relate to the observed emission structures.

In a forthcoming study, we will also degrade the spatial resolution of the NGC 4258 dataset to simulate how its morphology and emission-line structures would appear if the galaxy were observed at greater distances, such as that of NGC 7319 (D = 95.3 Mpc). This exercise will allow us to quantify how spatial resolution impacts our interpretation of AGN feedback morphologies—specifically, how discrete parsec-scale \feii and H$_2$ knots blend into more diffuse structures at lower resolution. Such modeling will help bridge the gap between nearby high-resolution case studies and more distant observations, providing critical context for interpreting JWST IFU data of AGN hosts where unresolved shock structures may mask the underlying complexity. While this analysis is beyond the scope of the current paper, it will be a key component of our upcoming comparative framework for connecting local and high-redshift AGN feedback signatures.

Finally, the geometry of several regions in NGC 4258 offers a unique opportunity to study energy stratification within the interstellar medium. Region B, in particular, shows compelling evidence of shocks interacting with PAH-emitting regions, possibly leading to PAH destruction. This is evident in the upper right part of the region, where H$_2$ emission appears immediately adjacent to PAH emission. Detailed analysis of these transition zones will allow us to quantify the energy deposition and its effects on different gas and dust components in the galaxy.

\section{Summary} 
\label{sec:summary}
In this paper, we presented an overview of JWST NIRCam imaging of NGC 4258, focusing on multiple near-infrared emission lines to trace shocks and study their effects on the ISM at parsec-scale resolution. We performed continuum subtraction to isolate key emission features, including \feii, H$_2$, Br$\alpha$, and PAH 3.3 $\mu$m, and compared these to archival UV, optical, radio, and X-ray imaging. Our analysis provides new insights into the interaction between AGN-driven winds and the surrounding ISM, revealing complex shock structures and stratified emission across several regions of the galaxy. 

\begin{enumerate}
\item Multi-phase Shock Diagnostics in NGC 4258:
   The observations reveal detailed evidence of multi-phase shocks interacting with the ISM. \feii and H$_2$ emission show distinct spatial correlations with the anomalous radio structure, indicating moderate-velocity shocks (50–300 km/s). The coexistence of \feii and H$_2$ suggests the presence of both J-type and C-type shocks, with high-velocity shocks destroying dust and liberating iron into the gas phase while lower-velocity shocks preserve molecular gas.  

\item AGN-driven Feedback and Its Impact on the ISM:
   The shock structures observed are consistent with AGN-driven winds interacting with dense regions of the galactic disk, particularly along the anomalous radio arc. Evidence of energy stratification is found in several regions, notably in Region B, where H$_2$ emission appears adjacent to PAH structures, possibly indicating PAH destruction by shock interactions.  

\item PAH Destruction by Shocks:  
   The absence of PAH emission in some shock-dominated regions suggests that high-velocity shocks ($v > 100$ km/s) are responsible for destroying PAH molecules. In contrast, PAH emission is preserved in regions associated with stellar winds and star formation, where shock velocities are likely lower.  

\item Utility of JWST for AGN Feedback Studies:
   JWST’s high resolution and sensitivity enable the detection of faint shock-excited emission features that were previously unobservable. These observations serve as a local benchmark for understanding AGN feedback in high-redshift galaxies and demonstrate the need for future NIRSpec and MIRI spectroscopy to fully characterize the kinematics and physical conditions of the ISM.  

\item Importance of Geometry and Projection Effects:
   The geometry of several regions in NGC 4258 provides a unique opportunity to study energy stratification within the ISM. Region B, in particular, offers compelling evidence of shocks interacting with PAH-emitting regions, highlighting the importance of projection effects in understanding the spatial distribution of different gas phases.  

\end{enumerate}

Together, these findings offer critical insights into the impact of AGN-driven feedback on the ISM in NGC 4258 and establish a foundation for future studies that will expand this analysis to more distant galaxies.


\section*{Acknowledgments} 
\label{sec:acknowledgements}
\noindent The authors thank the anonymous referee for their helpful comments that improved the clarity of this work. \\
TCF is thankful for the support of the European Space Agency (ESA). \\
NFC and ON were supported by NASA Postdoctoral Program Fellowships at NASA Goddard Space Flight Center, \\ 
administered by Oak Ridge Associated Universities under contract with NASA. \\
The authors thank J. DePasquale for providing the WCS-solved FITS image from the press image of M106. \\
The authors thank D. Michael Crenshaw, Julia Falcone, Brandon Hensley, Steve Kraemer, Beena Meena, \\ 
Mitchell Revalski, Maura Shea, and Krista Lynne Smith for helpful discussions while developing this manuscript. \\
This research also made use of Astropy, \footnote{\url{http://www.astropy.org}},\\  
a community-developed core Python package for Astronomy. \\
This work is based on observations made with the NASA/ESA/CSA James Webb Space Telescope. \\
The data were obtained from the Mikulski Archive for Space Telescopes at the Space Telescope Science Institute, \\ 
which is operated by the Association of Universities for Research in Astronomy, Inc., under NASA contract NAS 5-03127 for James Webb Space Telescope. \\
This research has made use of the NASA/IPAC Extragalactic Database (NED), which is funded by the \\ 
National Aeronautics and Space Administration and operated by the California Institute of Technology.



\section*{Data Availability} 
\label{sec:dataavilability}
The JWST data used in this study may be obtained from the Mikulski
Archive for Space Telescopes (\url{https://mast.stsci.edu/}) and
are associated with guest observer program \#2080. The specific observations analyzed can be accessed via \dataset[this DOI link]{http://dx.doi.org/10.17909/9xh7-n114}.


\bibliography{OverviewNGC4258_20240520}

\begin{thebibliography}{111}
\expandafter\ifx\csname natexlab\endcsname\relax\def\natexlab#1{#1}\fi

\bibitem[{{Aladro} {et~al.}(2013){Aladro}, {Viti}, {Bayet}, {Riquelme}, {Mart{\'\i}n}, {Mauersberger}, {Mart{\'\i}n-Pintado}, {Requena-Torres}, {Kramer}, \& {Wei{\ss}}}]{Ala13}
{Aladro}, R., {Viti}, S., {Bayet}, E., {et~al.} 2013, \aap, 549, A39

\bibitem[{{Allamandola} {et~al.}(1989){Allamandola}, {Tielens}, \& {Barker}}]{All89}
{Allamandola}, L.~J., {Tielens}, A.~G.~G.~M., \& {Barker}, J.~R. 1989, \apjs, 71, 733

\bibitem[{{Allen} {et~al.}(2008){Allen}, {Groves}, {Dopita}, {Sutherland}, \& {Kewley}}]{All08}
{Allen}, M.~G., {Groves}, B.~A., {Dopita}, M.~A., {Sutherland}, R.~S., \& {Kewley}, L.~J. 2008, \apjs, 178, 20

\bibitem[{{Alonso-Herrero} {et~al.}(1997){Alonso-Herrero}, {Rieke}, {Rieke}, \& {Ruiz}}]{Alo97}
{Alonso-Herrero}, A., {Rieke}, M.~J., {Rieke}, G.~H., \& {Ruiz}, M. 1997, \apj, 482, 747

\bibitem[{{Alonso-Herrero} {et~al.}(2014){Alonso-Herrero}, {Ramos Almeida}, {Esquej}, {Roche}, {Hern{\'a}n-Caballero}, {H{\"o}nig}, {Gonz{\'a}lez-Mart{\'\i}n}, {Aretxaga}, {Mason}, {Packham}, {Levenson}, {Rodr{\'\i}guez Espinosa}, {Siebenmorgen}, {Pereira-Santaella}, {D{\'\i}az-Santos}, {Colina}, {Alvarez}, \& {Telesco}}]{Alo14}
{Alonso-Herrero}, A., {Ramos Almeida}, C., {Esquej}, P., {et~al.} 2014, \mnras, 443, 2766

\bibitem[{{Appleton} {et~al.}(2006){Appleton}, {Xu}, {Reach}, {Dopita}, {Gao}, {Lu}, {Popescu}, {Sulentic}, {Tuffs}, \& {Yun}}]{App06}
{Appleton}, P.~N., {Xu}, K.~C., {Reach}, W., {et~al.} 2006, \apjl, 639, L51

\bibitem[{Appleton {et~al.}(2018)Appleton, Diaz-Santos, Fadda, Ogle, Togi, Lanz, Alatalo, Fischer, Rich, \& Guillard}]{appleton_jet-related_2018}
Appleton, P.~N., Diaz-Santos, T., Fadda, D., {et~al.} 2018, The Astrophysical Journal, 869, 61

\bibitem[{{Appleton} {et~al.}(2023){Appleton}, {Guillard}, {Emonts}, {Boulanger}, {Togi}, {Reach}, {Alatalo}, {Cluver}, {Diaz Santos}, {Duc}, {Gallagher}, {Ogle}, {O'Sullivan}, {Voggel}, \& {Xu}}]{App23}
{Appleton}, P.~N., {Guillard}, P., {Emonts}, B., {et~al.} 2023, \apj, 951, 104

\bibitem[{{Asada} \& {Nakamura}(2012)}]{Asa12}
{Asada}, K., \& {Nakamura}, M. 2012, \apjl, 745, L28

\bibitem[{{Barbosa} {et~al.}(2014){Barbosa}, {Storchi-Bergmann}, {McGregor}, {Vale}, \& {Rogemar Riffel}}]{Bar14}
{Barbosa}, F.~K.~B., {Storchi-Bergmann}, T., {McGregor}, P., {Vale}, T.~B., \& {Rogemar Riffel}, A. 2014, \mnras, 445, 2353

\bibitem[{{Black} \& {van Dishoeck}(1987)}]{Bla87}
{Black}, J.~H., \& {van Dishoeck}, E.~F. 1987, \apj, 322, 412

\bibitem[{Bushouse {et~al.}(2023)Bushouse, Eisenhamer, Dencheva, Davies, Greenfield, Morrison, Hodge, Simon, Grumm, Droettboom, Slavich, Sosey, Pauly, Miller, Jedrzejewski, Hack, Davis, Crawford, Law, Gordon, Regan, Cara, MacDonald, Bradley, Shanahan, Jamieson, Teodoro, Williams, \& Pena-Guerrero}]{bushouse_2023}
Bushouse, H., Eisenhamer, J., Dencheva, N., {et~al.} 2023, JWST Calibration Pipeline

\bibitem[{{Calzetti} {et~al.}(2015){Calzetti}, {Lee}, {Sabbi}, {Adamo}, {Smith}, {Andrews}, {Ubeda}, {Bright}, {Thilker}, {Aloisi}, {Brown}, {Chandar}, {Christian}, {Cignoni}, {Clayton}, {da Silva}, {de Mink}, {Dobbs}, {Elmegreen}, {Elmegreen}, {Evans}, {Fumagalli}, {Gallagher}, {Gouliermis}, {Grebel}, {Herrero}, {Hunter}, {Johnson}, {Kennicutt}, {Kim}, {Krumholz}, {Lennon}, {Levay}, {Martin}, {Nair}, {Nota}, {{\"O}stlin}, {Pellerin}, {Prieto}, {Regan}, {Ryon}, {Schaerer}, {Schiminovich}, {Tosi}, {Van Dyk}, {Walterbos}, {Whitmore}, \& {Wofford}}]{Cal15}
{Calzetti}, D., {Lee}, J.~C., {Sabbi}, E., {et~al.} 2015, \aj, 149, 51

\bibitem[{{Cecil} {et~al.}(2000){Cecil}, {Greenhill}, {DePree}, {Nagar}, {Wilson}, {Dopita}, {P{\'e}rez-Fournon}, {Argon}, \& {Moran}}]{Cec00}
{Cecil}, G., {Greenhill}, L.~J., {DePree}, C.~G., {et~al.} 2000, \apj, 536, 675

\bibitem[{{Chary} {et~al.}(2000){Chary}, {Becklin}, {Evans}, {Neugebauer}, {Scoville}, {Matthews}, \& {Ressler}}]{Cha00}
{Chary}, R., {Becklin}, E.~E., {Evans}, A.~S., {et~al.} 2000, \apj, 531, 756

\bibitem[{{Chastenet} {et~al.}(2023){Chastenet}, {Sutter}, {Sandstrom}, {Belfiore}, {Egorov}, {Larson}, {Leroy}, {Liu}, {Rosolowsky}, {Thilker}, {Watkins}, {Williams}, {Barnes}, {Bigiel}, {Boquien}, {Chevance}, {Dale}, {Kruijssen}, {Emsellem}, {Grasha}, {Groves}, {Hassani}, {Hughes}, {Kreckel}, {Meidt}, {Pan}, {Querejeta}, {Schinnerer}, \& {Whitcomb}}]{Cha23}
{Chastenet}, J., {Sutter}, J., {Sandstrom}, K., {et~al.} 2023, \apjl, 944, L12

\bibitem[{{Chastenet} {et~al.}(2024){Chastenet}, {De Looze}, {Rela{\~n}o}, {Dale}, {Williams}, {Bianchi}, {Xilouris}, {Baes}, {Bolatto}, {Boyer}, {Casasola}, {Clark}, {Fraternali}, {Fritz}, {Galliano}, {Glover}, {Gordon}, {Hirashita}, {Kennicutt}, {Nagamine}, {Kirchschlager}, {Klessen}, {Koch}, {Levy}, {McCallum}, {Madden}, {McLeod}, {Meidt}, {Mosenkov}, {Richie}, {Saintonge}, {Sandstrom}, {Schneider}, {Sivkova}, {Smith}, {Smith}, {van der Wel}, {Walch}, {Walter}, \& {Wood}}]{Cha24}
{Chastenet}, J., {De Looze}, I., {Rela{\~n}o}, M., {et~al.} 2024, \aap, 690, A348

\bibitem[{{Chen} {et~al.}(2024){Chen}, {Laor}, {Behar}, {Baldi}, {Gelfand}, {Kimball}, {McHardy}, {Orosz}, \& {Paragi}}]{Che24}
{Chen}, S., {Laor}, A., {Behar}, E., {et~al.} 2024, \apj, 975, 35

\bibitem[{{D'Agostino} {et~al.}(2019){D'Agostino}, {Kewley}, {Groves}, {Medling}, {Di Teodoro}, {Dopita}, {Thomas}, {Sutherland}, \& {Garcia-Burillo}}]{Dag19}
{D'Agostino}, J.~J., {Kewley}, L.~J., {Groves}, B.~A., {et~al.} 2019, \mnras, 487, 4153

\bibitem[{{Dasyra} {et~al.}(2024){Dasyra}, {Paraschos}, {Combes}, {Patapis}, {Helou}, {Papachristou}, {Fernandez-Ontiveros}, {Bisbas}, {Spinoglio}, {Armus}, \& {Malkan}}]{Das24}
{Dasyra}, K.~M., {Paraschos}, G.~F., {Combes}, F., {et~al.} 2024, \apj, 977, 156

\bibitem[{{Davies} {et~al.}(2024){Davies}, {Shimizu}, {Pereira-Santaella}, {Alonso-Herrero}, {Audibert}, {Bellocchi}, {Boorman}, {Campbell}, {Cao}, {Combes}, {Delaney}, {D{\'\i}az-Santos}, {Eisenhauer}, {Esparza Arredondo}, {Feuchtgruber}, {F{\"o}rster Schreiber}, {Fuller}, {Gandhi}, {Garc{\'\i}a-Bernete}, {Garc{\'\i}a-Burillo}, {Garc{\'\i}a-Lorenzo}, {Genzel}, {Gillessen}, {Gonz{\'a}lez Mart{\'\i}n}, {Haidar}, {Hermosa Mu{\~n}oz}, {Hicks}, {H{\"o}nig}, {Imanishi}, {Izumi}, {Labiano}, {Leist}, {Levenson}, {Lopez-Rodriguez}, {Lutz}, {Ott}, {Packham}, {Rabien}, {Ramos Almeida}, {Ricci}, {Rigopoulou}, {Rosario}, {Rouan}, {Santos}, {Shangguan}, {Stalevski}, {Sternberg}, {Sturm}, {Tacconi}, {Villar Mart{\'\i}n}, {Ward}, \& {Zhang}}]{Dav24}
{Davies}, R., {Shimizu}, T., {Pereira-Santaella}, M., {et~al.} 2024, \aap, 689, A263

\bibitem[{{Davies} {et~al.}(2007){Davies}, {M{\"u}ller S{\'a}nchez}, {Genzel}, {Tacconi}, {Hicks}, {Friedrich}, \& {Sternberg}}]{Dav07}
{Davies}, R.~I., {M{\"u}ller S{\'a}nchez}, F., {Genzel}, R., {et~al.} 2007, \apj, 671, 1388

\bibitem[{{Doi} {et~al.}(2013){Doi}, {Kohno}, {Nakanishi}, {Kameno}, {Inoue}, {Hada}, \& {Sorai}}]{Doi13}
{Doi}, A., {Kohno}, K., {Nakanishi}, K., {et~al.} 2013, \apj, 765, 63

\bibitem[{{Donnan} {et~al.}(2023){Donnan}, {Garc{\'\i}a-Bernete}, {Rigopoulou}, {Pereira-Santaella}, {Alonso-Herrero}, {Roche}, {Hern{\'a}n-Caballero}, \& {Spoon}}]{Don23}
{Donnan}, F.~R., {Garc{\'\i}a-Bernete}, I., {Rigopoulou}, D., {et~al.} 2023, \mnras, 519, 3691

\bibitem[{{Donnan} {et~al.}(2024){Donnan}, {Rigopoulou}, \& {Garc{\'\i}a-Bernete}}]{Don24}
{Donnan}, F.~R., {Rigopoulou}, D., \& {Garc{\'\i}a-Bernete}, I. 2024, \mnras, 532, L75

\bibitem[{{Dopita} {et~al.}(2005){Dopita}, {Groves}, {Fischera}, {Sutherland}, {Tuffs}, {Popescu}, {Kewley}, {Reuland}, \& {Leitherer}}]{Dop05}
{Dopita}, M.~A., {Groves}, B.~A., {Fischera}, J., {et~al.} 2005, \apj, 619, 755

\bibitem[{{Draine}(2011)}]{Dra11}
{Draine}, B.~T. 2011, {Physics of the Interstellar and Intergalactic Medium}

\bibitem[{{Draine} {et~al.}(2021){Draine}, {Li}, {Hensley}, {Hunt}, {Sandstrom}, \& {Smith}}]{Draine2021}
{Draine}, B.~T., {Li}, A., {Hensley}, B.~S., {et~al.} 2021, \apj, 917, 3

\bibitem[{{Falcone} {et~al.}(2024){Falcone}, {Crenshaw}, {Fischer}, {Meena}, {Revalski}, {Shea}, {Riffel}, {Chapman}, {Ferree}, {Tutterow}, \& {Davis}}]{Fal24}
{Falcone}, J., {Crenshaw}, D.~M., {Fischer}, T.~C., {et~al.} 2024, \apj, 971, 17

\bibitem[{{Fiore} {et~al.}(2017){Fiore}, {Feruglio}, {Shankar}, {Bischetti}, {Bongiorno}, {Brusa}, {Carniani}, {Cicone}, {Duras}, {Lamastra}, {Mainieri}, {Marconi}, {Menci}, {Maiolino}, {Piconcelli}, {Vietri}, \& {Zappacosta}}]{Fio17}
{Fiore}, F., {Feruglio}, C., {Shankar}, F., {et~al.} 2017, \aap, 601, A143

\bibitem[{{Fischer} {et~al.}(2023){Fischer}, {Johnson}, {Secrest}, {Crenshaw}, \& {Kraemer}}]{Fis23}
{Fischer}, T.~C., {Johnson}, M.~C., {Secrest}, N.~J., {Crenshaw}, D.~M., \& {Kraemer}, S.~B. 2023, \apj, 953, 87

\bibitem[{{Fischer} {et~al.}(2017){Fischer}, {Machuca}, {Diniz}, {Crenshaw}, {Kraemer}, {Riffel}, {Schmitt}, {Baron}, {Storchi-Bergmann}, {Straughn}, {Revalski}, \& {Pope}}]{Fis17}
{Fischer}, T.~C., {Machuca}, C., {Diniz}, M.~R., {et~al.} 2017, \apj, 834, 30

\bibitem[{{Fischer} {et~al.}(2018){Fischer}, {Kraemer}, {Schmitt}, {Longo Micchi}, {Crenshaw}, {Revalski}, {Vestergaard}, {Elvis}, {Gaskell}, {Hamann}, {Ho}, {Hutchings}, {Mushotzky}, {Netzer}, {Storchi-Bergmann}, {Straughn}, {Turner}, \& {Ward}}]{Fis18}
{Fischer}, T.~C., {Kraemer}, S.~B., {Schmitt}, H.~R., {et~al.} 2018, \apj, 856, 102

\bibitem[{{Fischer} {et~al.}(2021){Fischer}, {Secrest}, {Johnson}, {Dorland}, {Cigan}, {Fernandez}, {Hunt}, {Koss}, {Schmitt}, \& {Zacharias}}]{Fis21}
{Fischer}, T.~C., {Secrest}, N.~J., {Johnson}, M.~C., {et~al.} 2021, \apj, 906, 88

\bibitem[{{Flower} \& {Pineau Des For{\^e}ts}(2010)}]{Flo10}
{Flower}, D.~R., \& {Pineau Des For{\^e}ts}, G. 2010, \mnras, 406, 1745

\bibitem[{{Garc{\'\i}a-Bernete} {et~al.}(2022){Garc{\'\i}a-Bernete}, {Rigopoulou}, {Alonso-Herrero}, {Donnan}, {Roche}, {Pereira-Santaella}, {Labiano}, {Peralta de Arriba}, {Izumi}, {Ramos Almeida}, {Shimizu}, {H{\"o}nig}, {Garc{\'\i}a-Burillo}, {Rosario}, {Ward}, {Bellocchi}, {Hicks}, {Fuller}, \& {Packham}}]{Gar22}
{Garc{\'\i}a-Bernete}, I., {Rigopoulou}, D., {Alonso-Herrero}, A., {et~al.} 2022, \aap, 666, L5

\bibitem[{{Garc{\'\i}a-Bernete} {et~al.}(2024){Garc{\'\i}a-Bernete}, {Rigopoulou}, {Donnan}, {Alonso-Herrero}, {Pereira-Santaella}, {Shimizu}, {Davies}, {Roche}, {Garc{\'\i}a-Burillo}, {Labiano}, {Hermosa Mu{\~n}oz}, {Zhang}, {Audibert}, {Bellocchi}, {Bunker}, {Combes}, {Delaney}, {Esparza-Arredondo}, {Gandhi}, {Gonz{\'a}lez-Mart{\'\i}n}, {H{\"o}nig}, {Imanishi}, {Hicks}, {Fuller}, {Leist}, {Levenson}, {Lopez-Rodriguez}, {Packham}, {Ramos Almeida}, {Ricci}, {Stalevski}, {Villar Mart{\'\i}n}, \& {Ward}}]{Gar24}
{Garc{\'\i}a-Bernete}, I., {Rigopoulou}, D., {Donnan}, F.~R., {et~al.} 2024, \aap, 691, A162

\bibitem[{{Girdhar} {et~al.}(2022){Girdhar}, {Harrison}, {Mainieri}, {Bittner}, {Costa}, {Kharb}, {Mukherjee}, {Arrigoni Battaia}, {Alexander}, {Calistro Rivera}, {Circosta}, {De Breuck}, {Edge}, {Farina}, {Kakkad}, {Lansbury}, {Molyneux}, {Mullaney}, {Silpa}, {Thomson}, \& {Ward}}]{Gir22}
{Girdhar}, A., {Harrison}, C.~M., {Mainieri}, V., {et~al.} 2022, \mnras, 512, 1608

\bibitem[{{Guillard} {et~al.}(2009){Guillard}, {Boulanger}, {Pineau Des For{\^e}ts}, \& {Appleton}}]{Gui09}
{Guillard}, P., {Boulanger}, F., {Pineau Des For{\^e}ts}, G., \& {Appleton}, P.~N. 2009, \aap, 502, 515

\bibitem[{{Guillet} {et~al.}(2011){Guillet}, {Pineau Des For{\^e}ts}, \& {Jones}}]{Gui11}
{Guillet}, V., {Pineau Des For{\^e}ts}, G., \& {Jones}, A.~P. 2011, \aap, 527, A123

\bibitem[{{Harrison} \& {Ramos Almeida}(2024)}]{Har24}
{Harrison}, C.~M., \& {Ramos Almeida}, C. 2024, Galaxies, 12, 17

\bibitem[{{Hermosa Mu{\~n}oz} {et~al.}(2024){Hermosa Mu{\~n}oz}, {Alonso-Herrero}, {Pereira-Santaella}, {Garc{\'\i}a-Bernete}, {Garc{\'\i}a-Burillo}, {Garc{\'\i}a-Lorenzo}, {Davies}, {Shimizu}, {Esparza-Arredondo}, {Hicks}, {Haidar}, {Leist}, {L{\'o}pez-Rodr{\'\i}guez}, {Ramos Almeida}, {Rosario}, {Zhang}, {Audibert}, {Bellocchi}, {Boorman}, {Bunker}, {Combes}, {Campbell}, {D{\'\i}az-Santos}, {Fuller}, {Gandhi}, {Gonz{\'a}lez-Mart{\'\i}n}, {H{\"o}nig}, {Imanishi}, {Izumi}, {Labiano}, {Levenson}, {Packham}, {Ricci}, {Rigopoulou}, {Rouan}, {Stalevski}, {Villar-Mart{\'\i}n}, \& {Ward}}]{Her24}
{Hermosa Mu{\~n}oz}, L., {Alonso-Herrero}, A., {Pereira-Santaella}, M., {et~al.} 2024, \aap, 690, A350

\bibitem[{{Herrnstein} {et~al.}(2005){Herrnstein}, {Moran}, {Greenhill}, \& {Trotter}}]{Herrnstein_2005}
{Herrnstein}, J.~R., {Moran}, J.~M., {Greenhill}, L.~J., \& {Trotter}, A.~S. 2005, \apj, 629, 719

\bibitem[{{Herrnstein} {et~al.}(1999){Herrnstein}, {Moran}, {Greenhill}, {Diamond}, {Inoue}, {Nakai}, {Miyoshi}, {Henkel}, \& {Riess}}]{Her99}
{Herrnstein}, J.~R., {Moran}, J.~M., {Greenhill}, L.~J., {et~al.} 1999, \nat, 400, 539

\bibitem[{{Hollenbach} \& {McKee}(1989)}]{Hol89}
{Hollenbach}, D., \& {McKee}, C.~F. 1989, \apj, 342, 306

\bibitem[{{Hopkins} {et~al.}(2012){Hopkins}, {Quataert}, \& {Murray}}]{Hop12}
{Hopkins}, P.~F., {Quataert}, E., \& {Murray}, N. 2012, \mnras, 421, 3522

\bibitem[{{Jarvis} {et~al.}(2019){Jarvis}, {Harrison}, {Thomson}, {Circosta}, {Mainieri}, {Alexander}, {Edge}, {Lansbury}, {Molyneux}, \& {Mullaney}}]{Jar19}
{Jarvis}, M.~E., {Harrison}, C.~M., {Thomson}, A.~P., {et~al.} 2019, \mnras, 485, 2710

\bibitem[{{Kim} {et~al.}(2012){Kim}, {Im}, {Lee}, {Lee}, {Jun}, {Nakagawa}, {Matsuhara}, {Wada}, {Oyabu}, {Takagi}, {Inami}, {Ohyama}, {Yamada}, {Helou}, {Armus}, \& {Shi}}]{Kim12}
{Kim}, J.~H., {Im}, M., {Lee}, H.~M., {et~al.} 2012, \apj, 760, 120

\bibitem[{{Knop} {et~al.}(1996){Knop}, {Armus}, {Larkin}, {Mathews}, {Shupe}, \& {Soifer}}]{Kno96}
{Knop}, R.~A., {Armus}, L., {Larkin}, J.~E., {et~al.} 1996, \aj, 112, 81

\bibitem[{{Koo} \& {Lee}(2015)}]{Koo15}
{Koo}, B.-C., \& {Lee}, Y.-H. 2015, Publication of Korean Astronomical Society, 30, 145

\bibitem[{{Koo} {et~al.}(2016){Koo}, {Raymond}, \& {Kim}}]{Koo16}
{Koo}, B.-C., {Raymond}, J.~C., \& {Kim}, H.-J. 2016, Journal of Korean Astronomical Society, 49, 109

\bibitem[{Kristensen {et~al.}(2023)Kristensen, Godard, Guillard, Gusdorf, \& Pineau~des Forêts}]{Kri23}
Kristensen, L.~E., Godard, B., Guillard, P., Gusdorf, A., \& Pineau~des Forêts, G. 2023, Astronomy \& Astrophysics, 675

\bibitem[{{Lai} {et~al.}(2022){Lai}, {Armus}, {U}, {D{\'\i}az-Santos}, {Larson}, {Evans}, {Malkan}, {Appleton}, {Rich}, {M{\"u}ller-S{\'a}nchez}, {Inami}, {Bohn}, {McKinney}, {Finnerty}, {Law}, {Linden}, {Medling}, {Privon}, {Song}, {Stierwalt}, {van der Werf}, {Barcos-Mu{\~n}oz}, {Smith}, {Togi}, {Aalto}, {B{\"o}ker}, {Charmandaris}, {Howell}, {Iwasawa}, {Kemper}, {Mazzarella}, {Murphy}, {Brown}, {Hayward}, {Marshall}, {Sanders}, \& {Surace}}]{Lai22}
{Lai}, T. S.~Y., {Armus}, L., {U}, V., {et~al.} 2022, \apjl, 941, L36

\bibitem[{{Laine} {et~al.}(2010){Laine}, {Krause}, {Tabatabaei}, \& {Siopis}}]{Lai10}
{Laine}, S., {Krause}, M., {Tabatabaei}, F.~S., \& {Siopis}, C. 2010, \aj, 140, 1084

\bibitem[{{Launhardt} {et~al.}(2022){Launhardt}, {Loinard}, {Dzib}, {Forbrich}, {Bower}, {Henning}, {Mioduszewski}, \& {Reffert}}]{Lau22}
{Launhardt}, R., {Loinard}, L., {Dzib}, S.~A., {et~al.} 2022, \apj, 931, 43

\bibitem[{{Maloney} {et~al.}(1996){Maloney}, {Hollenbach}, \& {Tielens}}]{Mal96}
{Maloney}, P.~R., {Hollenbach}, D.~J., \& {Tielens}, A.~G.~G.~M. 1996, \apj, 466, 561

\bibitem[{{Maret} {et~al.}(2009){Maret}, {Bergin}, {Neufeld}, {Green}, {Watson}, {Harwit}, {Kristensen}, {Melnick}, {Sonnentrucker}, {Tolls}, {Werner}, {Willacy}, \& {Yuan}}]{Mar09}
{Maret}, S., {Bergin}, E.~A., {Neufeld}, D.~A., {et~al.} 2009, \apj, 698, 1244

\bibitem[{{Masini} {et~al.}(2022){Masini}, {Wijesekera}, {Celotti}, \& {Boorman}}]{Mas22}
{Masini}, A., {Wijesekera}, J.~V., {Celotti}, A., \& {Boorman}, P.~G. 2022, \aap, 663, A87

\bibitem[{{Mason} {et~al.}(2012){Mason}, {Lopez-Rodriguez}, {Packham}, {Alonso-Herrero}, {Levenson}, {Radomski}, {Ramos Almeida}, {Colina}, {Elitzur}, {Aretxaga}, {Roche}, \& {Oi}}]{Mas12}
{Mason}, R.~E., {Lopez-Rodriguez}, E., {Packham}, C., {et~al.} 2012, \aj, 144, 11

\bibitem[{{Meena} {et~al.}(2023){Meena}, {Crenshaw}, {Schmitt}, {Revalski}, {Chapman}, {Fischer}, {Kraemer}, {Robinson}, {Falcone}, \& {Polack}}]{Mee23}
{Meena}, B., {Crenshaw}, D.~M., {Schmitt}, H.~R., {et~al.} 2023, \apj, 943, 98

\bibitem[{{Meenakshi} {et~al.}(2022){Meenakshi}, {Mukherjee}, {Wagner}, {Nesvadba}, {Morganti}, {Janssen}, \& {Bicknell}}]{Mee22}
{Meenakshi}, M., {Mukherjee}, D., {Wagner}, A.~Y., {et~al.} 2022, \mnras, 511, 1622

\bibitem[{{Micelotta} {et~al.}(2010){Micelotta}, {Jones}, \& {Tielens}}]{Mic10}
{Micelotta}, E.~R., {Jones}, A.~P., \& {Tielens}, A.~G.~G.~M. 2010, \aap, 510, A36

\bibitem[{{Mukherjee} {et~al.}(2018){Mukherjee}, {Bicknell}, {Wagner}, {Sutherland}, \& {Silk}}]{Muk18}
{Mukherjee}, D., {Bicknell}, G.~V., {Wagner}, A.~Y., {Sutherland}, R.~S., \& {Silk}, J. 2018, \mnras, 479, 5544

\bibitem[{{Neufeld} {et~al.}(2006){Neufeld}, {Melnick}, {Sonnentrucker}, {Bergin}, {Green}, {Kim}, {Watson}, {Forrest}, \& {Pipher}}]{Neu06}
{Neufeld}, D.~A., {Melnick}, G.~J., {Sonnentrucker}, P., {et~al.} 2006, \apj, 649, 816

\bibitem[{{Nisini} {et~al.}(2002){Nisini}, {Caratti o Garatti}, {Giannini}, \& {Lorenzetti}}]{Nis02}
{Nisini}, B., {Caratti o Garatti}, A., {Giannini}, T., \& {Lorenzetti}, D. 2002, \aap, 393, 1035

\bibitem[{Ogle {et~al.}(2014)Ogle, Lanz, \& Appleton}]{ogle_jet-shocked_2014}
Ogle, P.~M., Lanz, L., \& Appleton, P.~N. 2014, The Astrophysical Journal, 788, L33

\bibitem[{{Oliva} {et~al.}(2001){Oliva}, {Marconi}, {Maiolino}, {Testi}, {Mannucci}, {Ghinassi}, {Licandro}, {Origlia}, {Baffa}, {Checcucci}, {Comoretto}, {Gavryussev}, {Gennari}, {Giani}, {Hunt}, {Lisi}, {Lorenzetti}, {Marcucci}, {Miglietta}, {Sozzi}, {Stefanini}, \& {Vitali}}]{Oli01}
{Oliva}, E., {Marconi}, A., {Maiolino}, R., {et~al.} 2001, \aap, 369, L5

\bibitem[{{Osterbrock} \& {Ferland}(2006)}]{Ost06}
{Osterbrock}, D.~E., \& {Ferland}, G.~J. 2006, {Astrophysics of gaseous nebulae and active galactic nuclei}

\bibitem[{{Otter} {et~al.}(2024){Otter}, {Alatalo}, {Rowlands}, {McDermid}, {Davis}, {Federrath}, {French}, {Heckman}, {Ogle}, {Kakkad}, {Luo}, {Nyland}, {Tripathi}, {Patil}, {Petric}, {Smercina}, {Skarbinski}, {Lanz}, {Larson}, {Appleton}, {Aalto}, {Olander}, {Sazonova}, \& {Smith}}]{Ott24}
{Otter}, J.~A., {Alatalo}, K., {Rowlands}, K., {et~al.} 2024, \apj, 975, 142

\bibitem[{{Panessa} {et~al.}(2019){Panessa}, {Baldi}, {Laor}, {Padovani}, {Behar}, \& {McHardy}}]{Pan19}
{Panessa}, F., {Baldi}, R.~D., {Laor}, A., {et~al.} 2019, Nature Astronomy, 3, 387

\bibitem[{{Peeters} {et~al.}(2004){Peeters}, {Spoon}, \& {Tielens}}]{Pee04}
{Peeters}, E., {Spoon}, H.~W.~W., \& {Tielens}, A.~G.~G.~M. 2004, \apj, 613, 986

\bibitem[{{Pereira-Santaella} {et~al.}(2022){Pereira-Santaella}, {{\'A}lvarez-M{\'a}rquez}, {Garc{\'\i}a-Bernete}, {Labiano}, {Colina}, {Alonso-Herrero}, {Bellocchi}, {Garc{\'\i}a-Burillo}, {H{\"o}nig}, {Ramos Almeida}, \& {Rosario}}]{Per22}
{Pereira-Santaella}, M., {{\'A}lvarez-M{\'a}rquez}, J., {Garc{\'\i}a-Bernete}, I., {et~al.} 2022, \aap, 665, L11

\bibitem[{{Polack} {et~al.}(2024){Polack}, {Revalski}, {Crenshaw}, {Fischer}, {Schmitt}, {Kraemer}, {Meena}, \& {Rafelski}}]{Pol24}
{Polack}, G.~E., {Revalski}, M., {Crenshaw}, D.~M., {et~al.} 2024, \apj, 975, 129

\bibitem[{{Ramos Almeida} {et~al.}(2022){Ramos Almeida}, {Bischetti}, {Garc{\'\i}a-Burillo}, {Alonso-Herrero}, {Audibert}, {Cicone}, {Feruglio}, {Tadhunter}, {Pierce}, {Pereira-Santaella}, \& {Bessiere}}]{Ram22}
{Ramos Almeida}, C., {Bischetti}, M., {Garc{\'\i}a-Burillo}, S., {et~al.} 2022, \aap, 658, A155

\bibitem[{Reid {et~al.}(2019)Reid, Pesce, \& Riess}]{Reid_2019}
Reid, M.~J., Pesce, D.~W., \& Riess, A.~G. 2019, The Astrophysical Journal Letters, 886, L27

\bibitem[{Rest(2024)}]{rest_jhat_2024}
Rest, A. 2024, JWST/HST Alignment Tool, \url{https://github.com/arminrest/jhat}

\bibitem[{{Richings} \& {Faucher-Gigu{\`e}re}(2018)}]{Ric18}
{Richings}, A.~J., \& {Faucher-Gigu{\`e}re}, C.-A. 2018, \mnras, 474, 3673

\bibitem[{Rieke {et~al.}(2009)Rieke, Alonso-Herrero, Weiner, Pérez-González, Blaylock, Donley, \& Marcillac}]{rieke_determining_2009}
Rieke, G.~H., Alonso-Herrero, A., Weiner, B.~J., {et~al.} 2009, The Astrophysical Journal, 692, 556

\bibitem[{Rieke \& Lebofsky(1985)}]{rieke_extinction_1985}
Rieke, G.~H., \& Lebofsky, M.~J. 1985, ApJ, 288, 618

\bibitem[{Rieke {et~al.}(2023)Rieke, Kelly, Misselt, Stansberry, Boyer, Beatty, Egami, Florian, Greene, Hainline, Leisenring, Roellig, Schlawin, Sun, Tinnin, Williams, Willmer, Wilson, Clark, Rohrbach, Brooks, Canipe, Correnti, DiFelice, Gennaro, Girard, Hartig, Hilbert, Koekemoer, Nikolov, Pirzkal, Rest, Robberto, Sunnquist, Telfer, Wu, Ferry, Lewis, Baum, Beichman, Doyon, Dressler, Eisenstein, Ferrarese, Hodapp, Horner, Jaffe, Johnstone, Krist, Martin, McCarthy, Meyer, Rieke, Trauger, \& Young}]{rieke_performance_2023}
Rieke, M.~J., Kelly, D.~M., Misselt, K., {et~al.} 2023, Publications of the Astronomical Society of the Pacific, 135, 028001, publisher: The Astronomical Society of the Pacific

\bibitem[{{Riffel} {et~al.}(2013){Riffel}, {Rodr{\'\i}guez-Ardila}, {Aleman}, {Brotherton}, {Pastoriza}, {Bonatto}, \& {Dors}}]{Rif13}
{Riffel}, R., {Rodr{\'\i}guez-Ardila}, A., {Aleman}, I., {et~al.} 2013, \mnras, 430, 2002

\bibitem[{Riffel {et~al.}(2014)Riffel, Vale, Storchi-Bergmann, \& McGregor}]{riffel_feeding_2014}
Riffel, R.~A., Vale, T.~B., Storchi-Bergmann, T., \& McGregor, P.~J. 2014, Monthly Notices of the Royal Astronomical Society, 442, 656

\bibitem[{{Rigopoulou} {et~al.}(2024){Rigopoulou}, {Donnan}, {Garc{\'\i}a-Bernete}, {Pereira-Santaella}, {Alonso-Herrero}, {Davies}, {Hunt}, {Roche}, \& {Shimizu}}]{Rig24}
{Rigopoulou}, D., {Donnan}, F.~R., {Garc{\'\i}a-Bernete}, I., {et~al.} 2024, \mnras, 532, 1598

\bibitem[{{Rodr{\'\i}guez} {et~al.}(2023){Rodr{\'\i}guez}, {Lee}, {Whitmore}, {Thilker}, {Maschmann}, {Chandar}, {Deger}, {Boquien}, {Dale}, {Larson}, {Williams}, {Kim}, {Schinnerer}, {Rosolowsky}, {Leroy}, {Emsellem}, {Sandstrom}, {Kruijssen}, {Grasha}, {Watkins}, {Barnes}, {Sormani}, {Kim}, {Anand}, {Chevance}, {Bigiel}, {Klessen}, {Hassani}, {Liu}, {Faesi}, {Cao}, {Belfiore}, {Pessa}, {Kreckel}, {Groves}, {Pety}, {Indebetouw}, {Egorov}, {Blanc}, {Saito}, \& {Hughes}}]{Rod23}
{Rodr{\'\i}guez}, M.~J., {Lee}, J.~C., {Whitmore}, B.~C., {et~al.} 2023, \apjl, 944, L26

\bibitem[{{Rodr{\'\i}guez-Ardila} {et~al.}(2004){Rodr{\'\i}guez-Ardila}, {Pastoriza}, {Viegas}, {Sigut}, \& {Pradhan}}]{Rod04}
{Rodr{\'\i}guez-Ardila}, A., {Pastoriza}, M.~G., {Viegas}, S., {Sigut}, T.~A.~A., \& {Pradhan}, A.~K. 2004, \aap, 425, 457

\bibitem[{{Rodr{\'\i}guez-Ardila} {et~al.}(2005){Rodr{\'\i}guez-Ardila}, {Riffel}, \& {Pastoriza}}]{Rod05}
{Rodr{\'\i}guez-Ardila}, A., {Riffel}, R., \& {Pastoriza}, M.~G. 2005, \mnras, 364, 1041

\bibitem[{{Saglia} {et~al.}(2016){Saglia}, {Opitsch}, {Erwin}, {Thomas}, {Beifiori}, {Fabricius}, {Mazzalay}, {Nowak}, {Rusli}, \& {Bender}}]{Sag16}
{Saglia}, R.~P., {Opitsch}, M., {Erwin}, P., {et~al.} 2016, \apj, 818, 47

\bibitem[{{Sandstrom} {et~al.}(2023){Sandstrom}, {Koch}, {Leroy}, {Rosolowsky}, {Emsellem}, {Smith}, {Egorov}, {Williams}, {Larson}, {Lee}, {Schinnerer}, {Thilker}, {Barnes}, {Belfiore}, {Bigiel}, {Blanc}, {Bolatto}, {Boquien}, {Cao}, {Chastenet}, {Chevance}, {Chiang}, {Dale}, {Faesi}, {Glover}, {Grasha}, {Groves}, {Hassani}, {Henshaw}, {Hughes}, {Kim}, {Klessen}, {Kreckel}, {Kruijssen}, {Lopez}, {Liu}, {Meidt}, {Murphy}, {Pan}, {Querejeta}, {Saito}, {Sardone}, {Sormani}, {Sutter}, {Usero}, \& {Watkins}}]{San23}
{Sandstrom}, K.~M., {Koch}, E.~W., {Leroy}, A.~K., {et~al.} 2023, \apjl, 944, L8

\bibitem[{{Sawada-Satoh} {et~al.}(2007){Sawada-Satoh}, {Ho}, {Muller}, {Matsushita}, \& {Lim}}]{Saw07}
{Sawada-Satoh}, S., {Ho}, P.~T.~P., {Muller}, S., {Matsushita}, S., \& {Lim}, J. 2007, \apj, 658, 851

\bibitem[{{Schmitt} {et~al.}(2003){Schmitt}, {Donley}, {Antonucci}, {Hutchings}, \& {Kinney}}]{Sch03}
{Schmitt}, H.~R., {Donley}, J.~L., {Antonucci}, R.~R.~J., {Hutchings}, J.~B., \& {Kinney}, A.~L. 2003, \apjs, 148, 327

\bibitem[{{Silk}(2013)}]{Sil13}
{Silk}, J. 2013, \apj, 772, 112

\bibitem[{{Sivasankaran} {et~al.}(2025){Sivasankaran}, {Blecha}, {Torrey}, {Kelley}, {Bhowmick}, {Vogelsberger}, {Hernquist}, {Marinacci}, \& {Sales}}]{Siv25}
{Sivasankaran}, A., {Blecha}, L., {Torrey}, P., {et~al.} 2025, \mnras, 537, 817

\bibitem[{{Smith} {et~al.}(2020){Smith}, {Koss}, {Mushotzky}, {Wong}, {Shimizu}, {Ricci}, \& {Ricci}}]{Smi20}
{Smith}, K.~L., {Koss}, M., {Mushotzky}, R., {et~al.} 2020, \apj, 904, 83

\bibitem[{{Sotira} {et~al.}(2024){Sotira}, {Vazza}, \& {Brighenti}}]{Sot24}
{Sotira}, S., {Vazza}, F., \& {Brighenti}, F. 2024, arXiv e-prints, arXiv:2410.07314

\bibitem[{Storchi-Bergmann {et~al.}(2009)Storchi-Bergmann, McGregor, Riffel, Simões~Lopes, Beck, \& Dopita}]{storchi-bergmann_feeding_2009}
Storchi-Bergmann, T., McGregor, P.~J., Riffel, R.~A., {et~al.} 2009, Monthly Notices of the Royal Astronomical Society, 394, 1148

\bibitem[{{Storchi-Bergmann} {et~al.}(2018){Storchi-Bergmann}, {Dall'Agnol de Oliveira}, {Longo Micchi}, {Schmitt}, {Fischer}, {Kraemer}, {Crenshaw}, {Maksym}, {Elvis}, {Fabbiano}, \& {Colina}}]{Sto18}
{Storchi-Bergmann}, T., {Dall'Agnol de Oliveira}, B., {Longo Micchi}, L.~F., {et~al.} 2018, \apj, 868, 14

\bibitem[{{Su} {et~al.}(2010){Su}, {Slatyer}, \& {Finkbeiner}}]{Su10}
{Su}, M., {Slatyer}, T.~R., \& {Finkbeiner}, D.~P. 2010, \apj, 724, 1044

\bibitem[{{Sutter} {et~al.}(2024){Sutter}, {Sandstrom}, {Chastenet}, {Leroy}, {Koch}, {Williams}, {Chown}, {Belfiore}, {Bigiel}, {Boquien}, {Cao}, {Chevance}, {Dale}, {Egorov}, {Glover}, {Groves}, {Klessen}, {Kreckel}, {Larson}, {Oakes}, {Pathak}, {Ramambason}, {Rosolowsky}, \& {Watkins}}]{Sut24}
{Sutter}, J., {Sandstrom}, K., {Chastenet}, J., {et~al.} 2024, \apj, 971, 178

\bibitem[{{Turner} \& {Ho}(1994)}]{Tur94}
{Turner}, J.~L., \& {Ho}, P. T.~P. 1994, \apj, 421, 122

\bibitem[{{Ujjwal} {et~al.}(2024){Ujjwal}, {Kartha}, {Akhil}, {Mathew}, {Subramanian}, {Sudheesh}, \& {Thomas}}]{Ujj24}
{Ujjwal}, K., {Kartha}, S.~S., {Akhil}, K.~R., {et~al.} 2024, \aap, 684, A71

\bibitem[{{van der Hulst} {et~al.}(1983){van der Hulst}, {Hummel}, {Davies}, {Pedlar}, \& {van Albada}}]{van83b}
{van der Hulst}, J.~M., {Hummel}, E., {Davies}, R.~D., {Pedlar}, A., \& {van Albada}, G.~D. 1983, \nat, 306, 566

\bibitem[{{van der Werf} {et~al.}(1993){van der Werf}, {Genzel}, {Krabbe}, {Blietz}, {Lutz}, {Drapatz}, {Ward}, \& {Forbes}}]{van93}
{van der Werf}, P.~P., {Genzel}, R., {Krabbe}, A., {et~al.} 1993, \apj, 405, 522

\bibitem[{{van Diedenhoven} {et~al.}(2004){van Diedenhoven}, {Peeters}, {Van Kerckhoven}, {Hony}, {Hudgins}, {Allamandola}, \& {Tielens}}]{vanD04}
{van Diedenhoven}, B., {Peeters}, E., {Van Kerckhoven}, C., {et~al.} 2004, \apj, 611, 928

\bibitem[{{Venturi} {et~al.}(2021){Venturi}, {Cresci}, {Marconi}, {Mingozzi}, {Nardini}, {Carniani}, {Mannucci}, {Marasco}, {Maiolino}, {Perna}, {Treister}, {Bland-Hawthorn}, \& {Gallimore}}]{Ven21}
{Venturi}, G., {Cresci}, G., {Marconi}, A., {et~al.} 2021, \aap, 648, A17

\bibitem[{{V{\'e}ron-Cetty} \& {V{\'e}ron}(2006)}]{Veron_2006}
{V{\'e}ron-Cetty}, M.~P., \& {V{\'e}ron}, P. 2006, \aap, 455, 773

\bibitem[{{Viti} {et~al.}(2014){Viti}, {Garc{\'\i}a-Burillo}, {Fuente}, {Hunt}, {Usero}, {Henkel}, {Eckart}, {Martin}, {Spaans}, {Muller}, {Combes}, {Krips}, {Schinnerer}, {Casasola}, {Costagliola}, {Marquez}, {Planesas}, {van der Werf}, {Aalto}, {Baker}, {Boone}, \& {Tacconi}}]{Vit14}
{Viti}, S., {Garc{\'\i}a-Burillo}, S., {Fuente}, A., {et~al.} 2014, \aap, 570, A28

\bibitem[{{Wagner} {et~al.}(2012){Wagner}, {Bicknell}, \& {Umemura}}]{Wag12}
{Wagner}, A.~Y., {Bicknell}, G.~V., \& {Umemura}, M. 2012, \apj, 757, 136

\bibitem[{Willott(2024)}]{willott_image1overf_2024}
Willott, C. 2024, jwst - Tools for processing and analyzing JWST data, \url{https://github.com/chriswillott/jwst}

\bibitem[{{Wolfire} {et~al.}(2022){Wolfire}, {Vallini}, \& {Chevance}}]{Wol22}
{Wolfire}, M.~G., {Vallini}, L., \& {Chevance}, M. 2022, \araa, 60, 247

\bibitem[{{Yuan} \& {Narayan}(2014)}]{Yua14}
{Yuan}, F., \& {Narayan}, R. 2014, \araa, 52, 529

\bibitem[{{Zubovas} {et~al.}(2013){Zubovas}, {Nayakshin}, {King}, \& {Wilkinson}}]{Zub13}
{Zubovas}, K., {Nayakshin}, S., {King}, A., \& {Wilkinson}, M. 2013, \mnras, 433, 3079

\end{thebibliography}

\appendix

\section{Continuum Subtraction Performance in Representative Regions}\label{section: cont_sub_ex}

\begin{figure*}
    \centering
    \includegraphics[width=\textwidth]{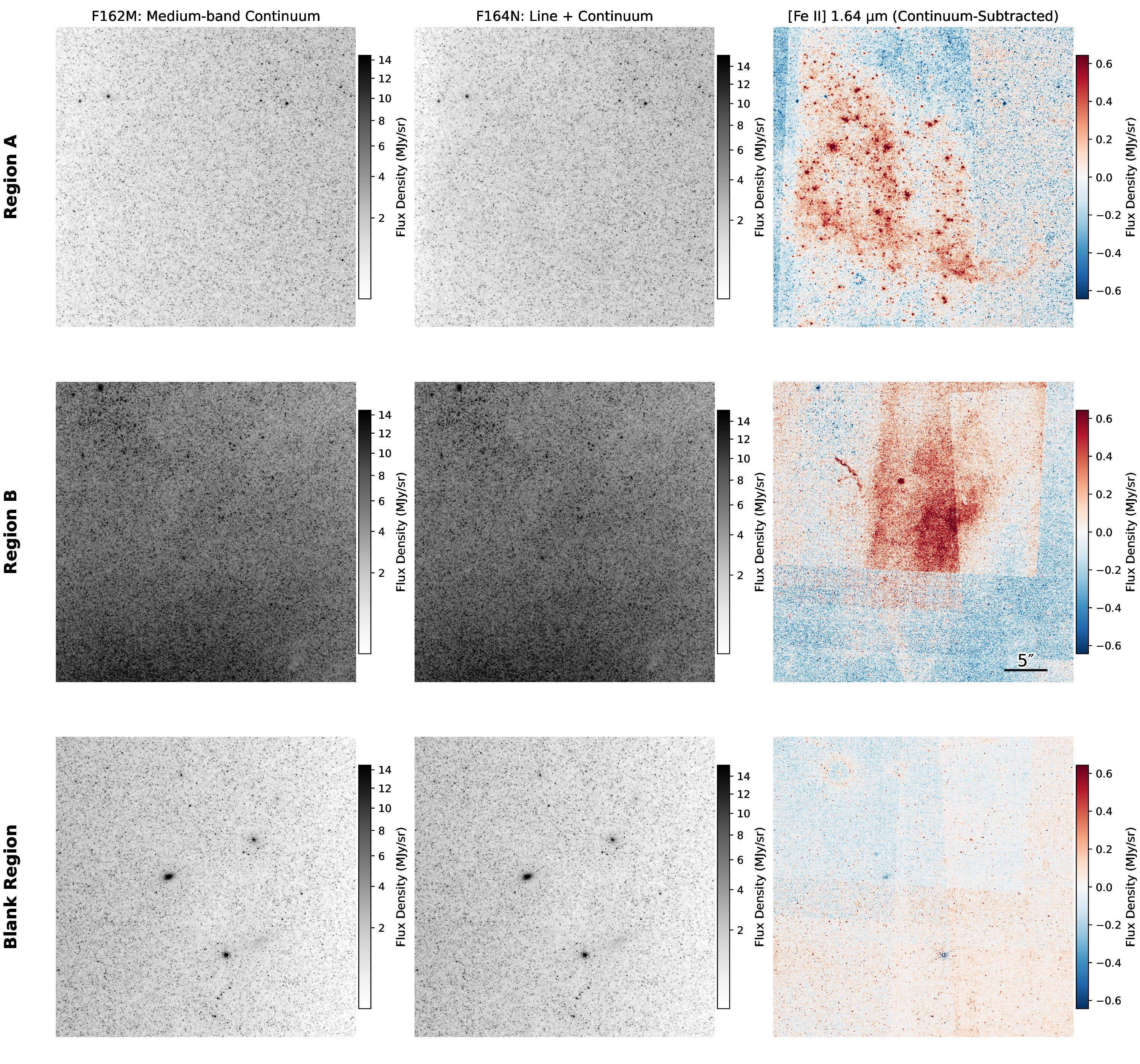}
    \caption{Each row shows a $35'' \times 35''$ cutout of a region selected from the NGC 4258 field: Region A (top) and Region B (middle) contain bright \feii emission arcs, while a select empty region (bottom) is devoid of obvious line emission. From left to right, columns show the medium-band continuum image (F162M), the narrow-band line+continuum image (F164N), and the continuum-subtracted \feii emission. Cutouts are oriented with North up and East to the left. Color scales for \feii are symmetric about zero to highlight both positive and negative residuals. A scale bar of 5$''$ is shown in the final column. The results demonstrate successful subtraction of continuum across both emission-dominated and line-free regions.}
    
    \label{fig:appendix_feii}
\end{figure*}

\begin{figure*}
    \centering
    \includegraphics[width=\textwidth]{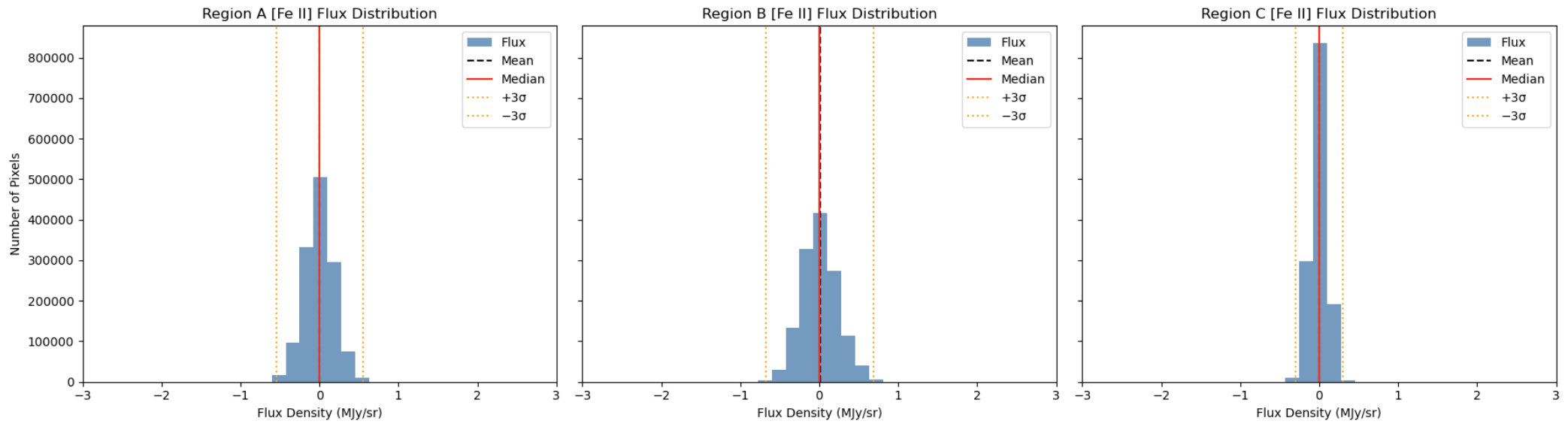}
    \caption{Histograms show the flux density distribution of \feii emission after continuum subtraction in Regions A, B, and C. Vertical lines mark the mean (black dashed), median (red solid), and $\pm3\sigma$ (orange dotted) levels. The distributions are sharply peaked and symmetric about zero, including Region C, which lacks line emission. The measured $1\sigma$ residual scatter in Region C (0.10\,MJy\,sr$^{-1}$) provides an estimate of the systematic uncertainty from the subtraction process.}
    
    \label{fig:appendix_bars}
\end{figure*}

To illustrate the effectiveness of our continuum subtraction method, we present \feii flux distributions for three representative $35'' \times 35''$ regions: Region A and Region B, which contain bright \feii arcs, and a third, blank-sky region with no obvious emission in Figure \ref{fig:appendix_feii}. For each region, we show histograms of pixel flux values, along with overplotted lines indicating the mean, median, and $\pm3\sigma$ levels in Figure \ref{fig:appendix_bars}. These distributions demonstrate that the continuum subtraction yields symmetric, narrow residuals centered near zero, even in the blank field region, which is most susceptible to sky background variation and mosaicing artifacts. The $1\sigma$ scatter in this blank region is 0.10\,MJy\,sr$^{-1}$, corresponding to a $\sim$4\,Jy uncertainty over the full aperture, consistent with the \feii flux errors reported in Table~\ref{tab:table1}. This analysis confirms that our subtraction method preserves faint emission while minimizing spurious residuals.

\clearpage 

\section{Region Image Alignments} \label{section: alignment}

To enhance the visualization of multi-wavelength observations in NGC 4258, we include an animated sequence illustrating spatial variations in key emission features across each analyzed Region shown in Figures \ref{fig:grid_nucleus}-\ref{fig:grid_E}. These figures provide a dynamic representation of PAH 3.3$\mu$m, H$_2$ 2.12$\mu$m, \feii 1.64$\mu$m, Br$\alpha$ 4.05$\mu$m and HST optical, along with overlaid radio continuum contours. The animation allows for a more intuitive comparison of structural differences across wavebands, highlighting the interaction between AGN-driven feedback and the surrounding interstellar medium. While each frame can be examined individually, the animation conveys the spatial relationships more effectively than static images alone. The inclusion of these figures in the manuscript is intended to provide a clearer, more comprehensive view of the multi-phase nature of the observed region.

\null
\vfill 

\begin{figure}[ht]
    \centering
    \includegraphics[width=0.7\linewidth]{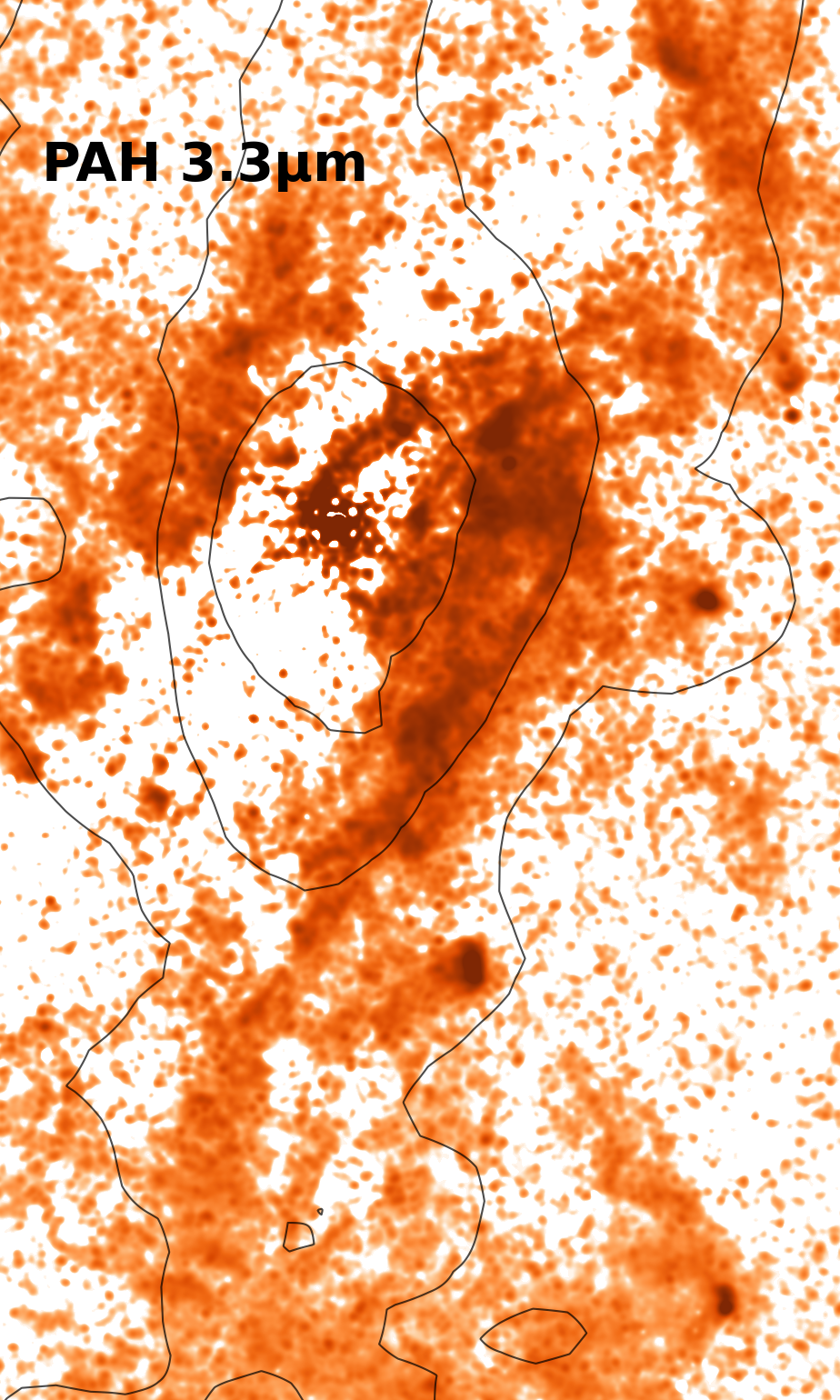}
    \caption{
    Static frame from an animated multi-wavelength sequence of Region N in NGC 4258 (available in the HTML version of the article). 
    The animation contains five frames spanning $\sim$2.5 s, sequentially displaying PAH 3.3$\mu$m (orange), H$_2$ 2.12$\mu$m (green), [Fe II] 1.64$\mu$m (blue), Br$\alpha$ 4.05$\mu$m (red), LEGUS F275W ultraviolet imaging (purple), and HST optical (grayscale). 
    }
    \label{fig:ngc4258_animationE}
\end{figure}

\vfill 

\clearpage

\null
\vfill 

\begin{figure}[ht]
    \centering
    \includegraphics[width=0.8\linewidth]{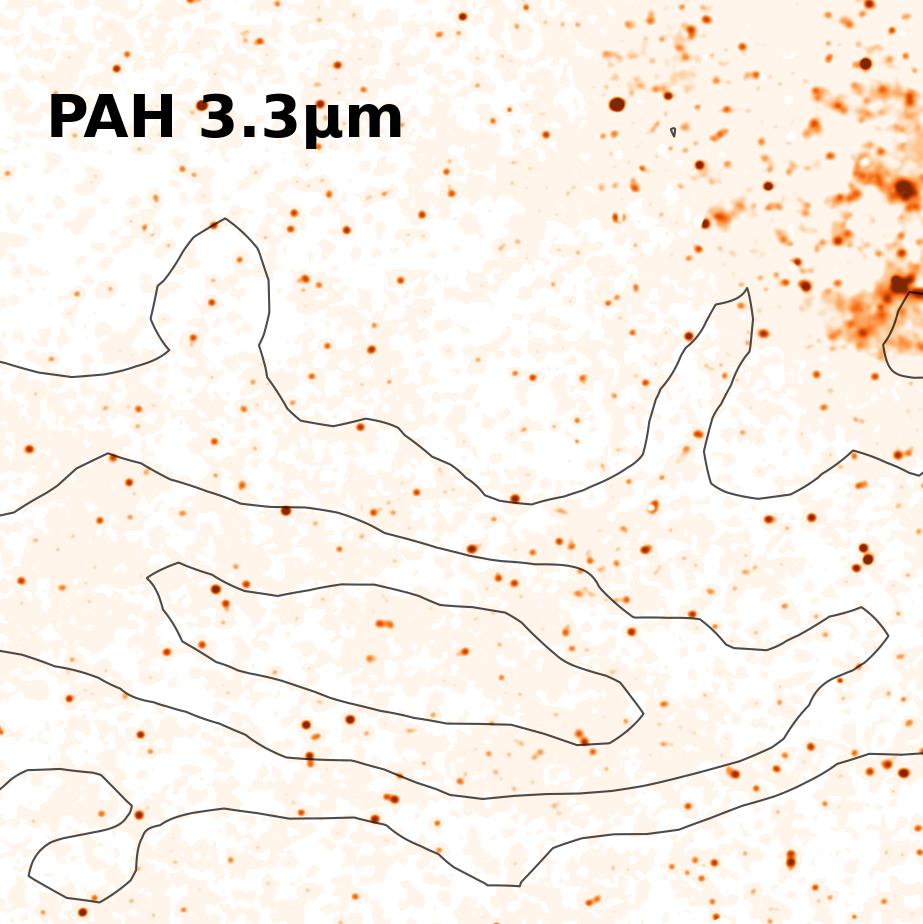}
    \caption{
    Static frame from an animated multi-wavelength sequence of Region A in NGC 4258 (available in the HTML version of the article). 
    The animation cycles through five frames over $\sim$2.5 seconds (0.5 s per frame), showing PAH 3.3$\mu$m (orange), H$_2$ 2.12$\mu$m (green), [Fe II] 1.64$\mu$m (blue), Br$\alpha$ 4.05$\mu$m (red), LEGUS F275W ultraviolet imaging (purple), and HST optical (grayscale). 
    Contours represent 4.89 GHz radio continuum emission. 
    }
    \label{fig:ngc4258_animationA}
\end{figure}

\vfill 

\clearpage

\null
\vfill 

\begin{figure}[ht]
    \centering
    \includegraphics[width=0.8\linewidth]{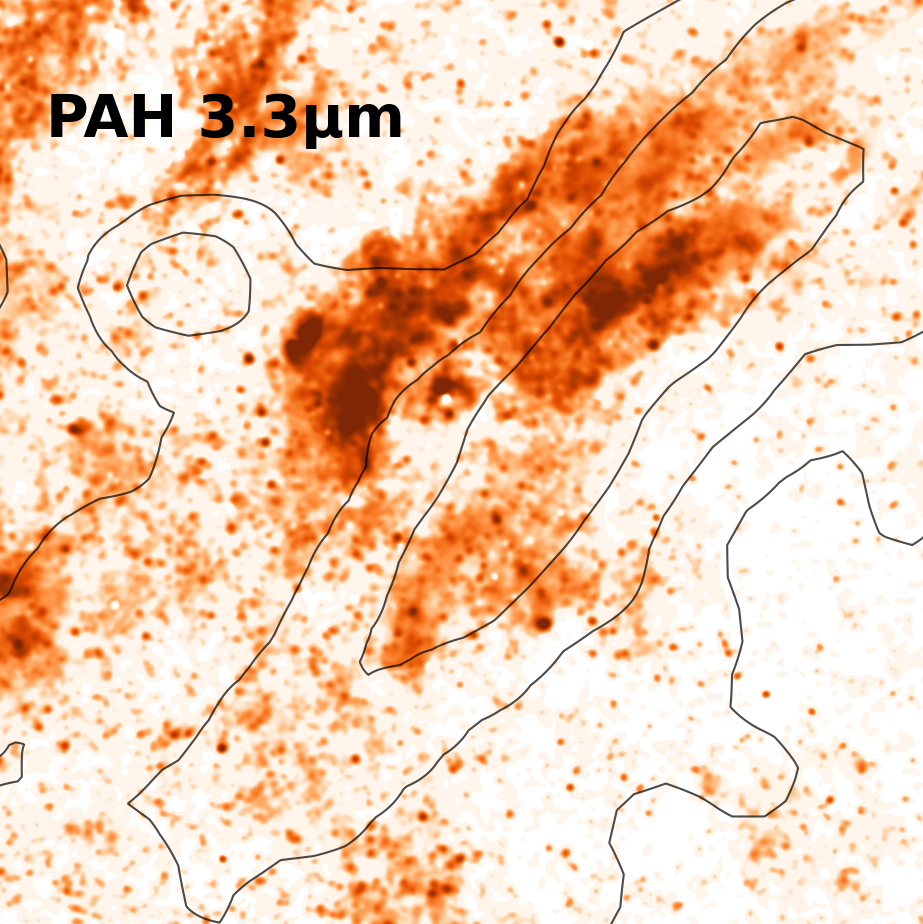}
    \caption{
    Static frame from an animated multi-wavelength sequence of Region B in NGC 4258 (available in the HTML version of the article). 
    The five-frame animation ($\sim$2.5 s total duration) alternates between PAH 3.3$\mu$m (orange), H$_2$ 2.12$\mu$m (green), [Fe II] 1.64$\mu$m (blue), Br$\alpha$ 4.05$\mu$m (red), LEGUS F275W ultraviolet imaging (purple), and HST optical (grayscale). 
    Radio continuum contours are overlaid at 4.89 GHz. 
    }
    \label{fig:ngc4258_animationB}
\end{figure}

\vfill 

\clearpage

\null
\vfill 

\begin{figure}[ht]
    \centering
    \includegraphics[width=0.8\linewidth]{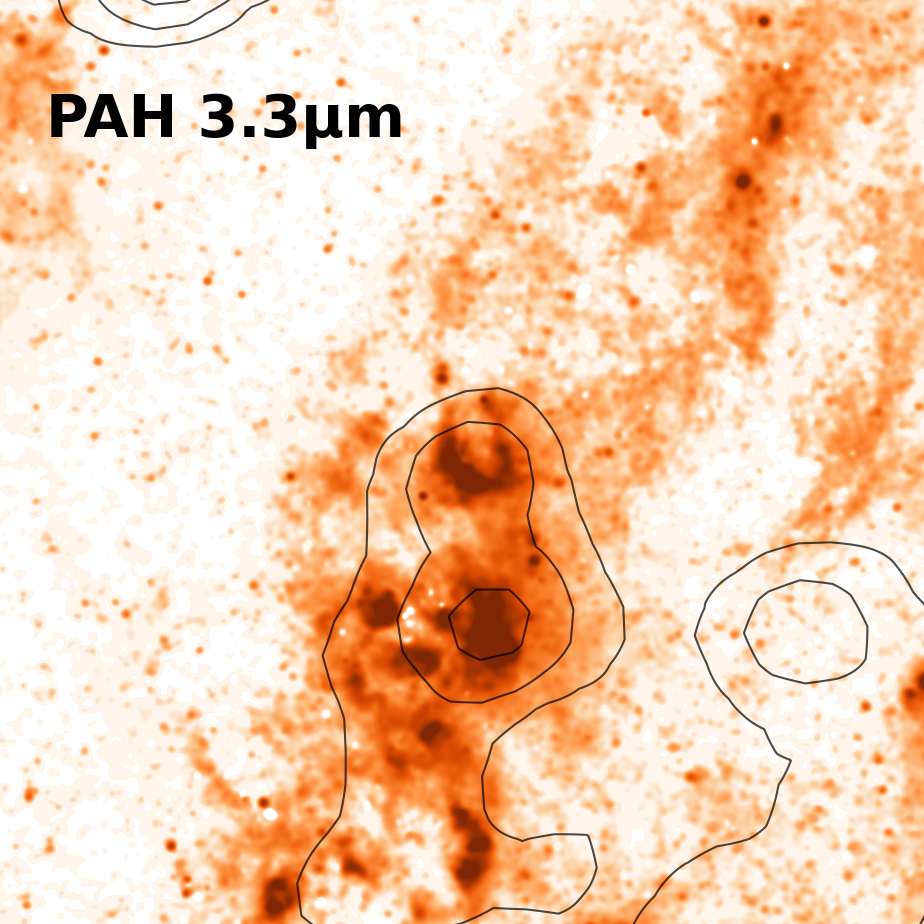}
    \caption{
    Static frame from an animated multi-wavelength sequence of Region C in NGC 4258 (available in the HTML version of the article). 
    The animation contains five frames displayed over $\sim$2.5 s, highlighting PAH 3.3$\mu$m (orange), H$_2$ 2.12$\mu$m (green), [Fe II] 1.64$\mu$m (blue), Br$\alpha$ 4.05$\mu$m (red), LEGUS F275W ultraviolet imaging (purple), and HST optical (grayscale). 
    Radio continuum contours trace 4.89 GHz emission. 
    }
    \label{fig:ngc4258_animationC}
\end{figure}

\vfill 

\clearpage

\null
\vfill 

\begin{figure}[ht]
    \centering
    \includegraphics[width=0.8\linewidth]{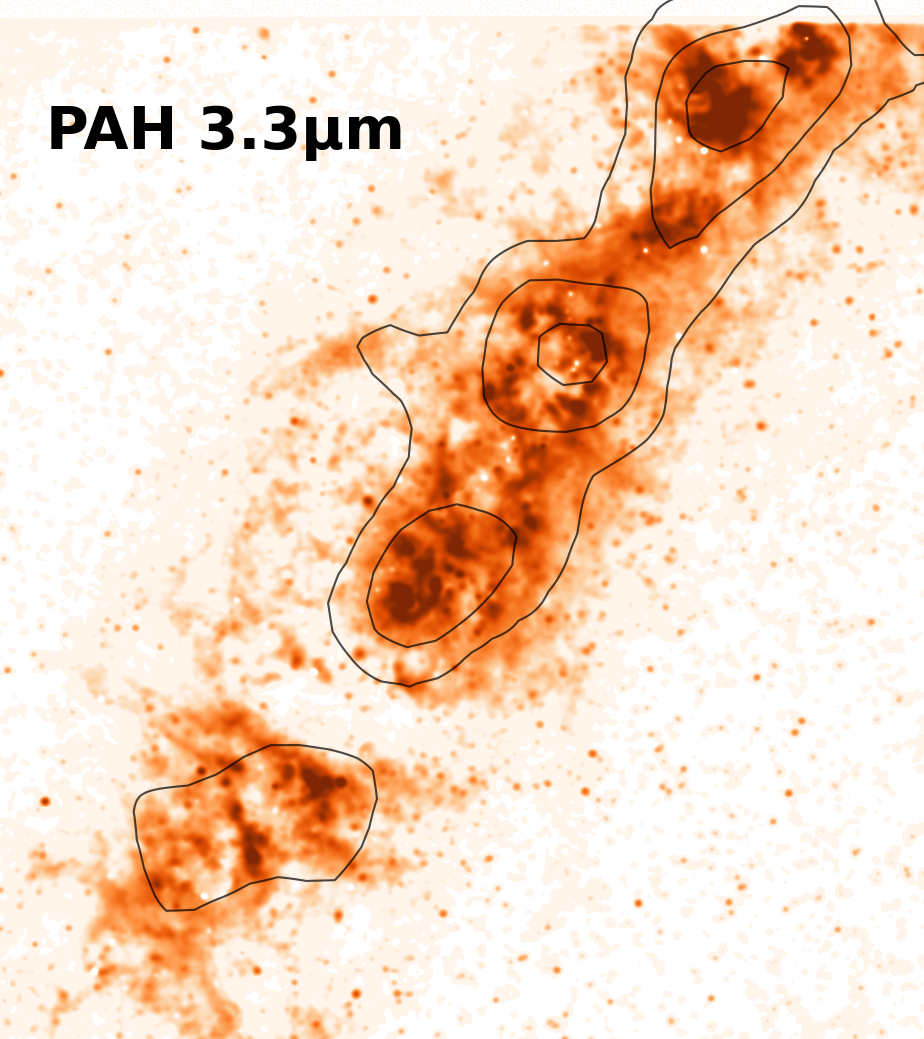}
    \caption{
    Static frame from an animated multi-wavelength sequence of Region D in NGC 4258 (available in the HTML version of the article). 
    The five-frame animation ($\sim$2.5 s) alternates between PAH 3.3$\mu$m (orange), H$_2$ 2.12$\mu$m (green), [Fe II] 1.64$\mu$m (blue), Br$\alpha$ 4.05$\mu$m (red), LEGUS F275W ultraviolet imaging (purple), and HST optical (grayscale). 
    Contours show 4.89 GHz radio continuum emission. 
    }
    \label{fig:ngc4258_animationD}
\end{figure}

\vfill 

\clearpage

\null
\vfill 

\begin{figure}[ht]
    \centering
    \includegraphics[width=0.8\linewidth]{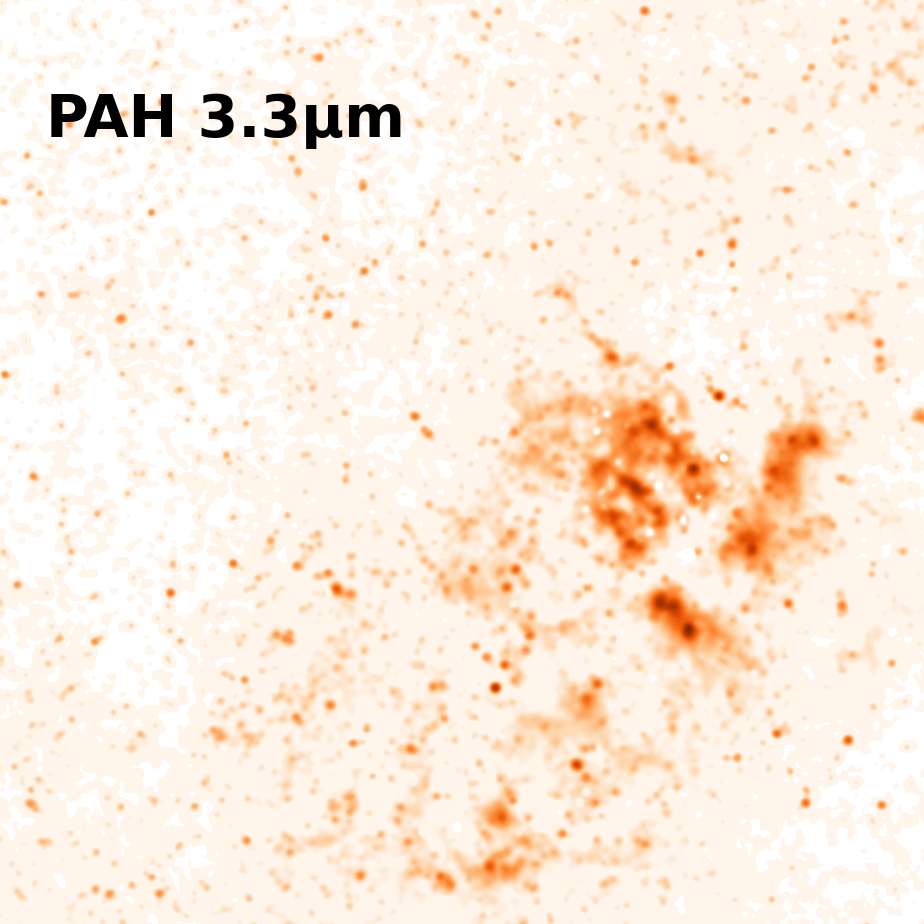}
    \caption{
    Static frame from an animated multi-wavelength sequence of Region E in NGC 4258 (available in the HTML version of the article). 
    The animation contains five frames spanning $\sim$2.5 s, sequentially displaying PAH 3.3$\mu$m (orange), H$_2$ 2.12$\mu$m (green), [Fe II] 1.64$\mu$m (blue), Br$\alpha$ 4.05$\mu$m (red), LEGUS F275W ultraviolet imaging (purple), and HST optical (grayscale). 
    }
    \label{fig:ngc4258_animationE}
\end{figure}

\vfill 

\end{document}